\documentclass[journal]{IEEEtran}
\IEEEoverridecommandlockouts
\usepackage{tipa}
\usepackage{bbm}
\usepackage{mathrsfs}
\usepackage{cite,url,subfigure,epsfig,graphicx}
\usepackage{amssymb,amsmath}
\usepackage{indentfirst}
\usepackage{pdfpages}
\usepackage{times,verbatim,amsfonts,amsmath,color}
\usepackage{cases}
\usepackage{bm}
\usepackage[ruled,vlined]{algorithm2e}
\usepackage{lineno}

\begin{document}
%\linenumbers

\title{Social-Aware Clustered Federated Learning with Customized Privacy Preservation}
\author{
\IEEEauthorblockN{Yuntao~Wang\IEEEauthorrefmark{2}, Zhou~Su\IEEEauthorrefmark{2}\IEEEauthorrefmark{1}, Yanghe Pan\IEEEauthorrefmark{2}, Tom H Luan\IEEEauthorrefmark{2}, Ruidong~Li\IEEEauthorrefmark{4}, and Shui Yu\IEEEauthorrefmark{5}}\\
\IEEEauthorblockA{
\IEEEauthorrefmark{2}School of Cyber Science and Engineering, Xi'an Jiaotong University, China\\
\IEEEauthorrefmark{4}Institute of Science and Engineering, Kanazawa University, Japan\\
\IEEEauthorrefmark{5}School of Computer Science, University of Technology Sydney, Australia
}
\thanks{This paper has been accepted by IEEE/ACM Transactions on Networking in March 2024.}}
\maketitle

\begin{abstract}
A key feature of federated learning (FL) is to preserve the data privacy of end users. However, there still exist potential privacy leakage in exchanging gradients under FL. As a result, recent research often explores the differential privacy (DP) approaches to add noises to the computing results to address privacy concerns with low overheads, which however degrade the model performance. In this paper, we strike the balance of data privacy and efficiency by utilizing the pervasive social connections between users.
Specifically, we propose SCFL, a novel  \underline{\textbf{S}}ocial-aware \underline{\textbf{C}}lustered \underline{\textbf{F}}ederated \underline{\textbf{L}}earning scheme, where mutually trusted individuals can freely form a social cluster and aggregate their raw model updates (e.g., gradients) inside each cluster before uploading to the cloud for global aggregation. By mixing model updates in a social group, adversaries can only eavesdrop the social-layer combined results, but not the privacy of individuals. As such, SCFL considerably enhances model utility without sacrificing privacy in a low-cost and highly feasible manner. We unfold the design of SCFL in three steps. \emph{i)~Stable social cluster formation}. Considering users' heterogeneous training samples and data distributions, we formulate the optimal social cluster formation problem as a federation game and devise a fair revenue allocation mechanism to resist free-riders. \emph{ii)~Differentiated trust-privacy mapping}. For the clusters with low mutual trust, we design a customizable privacy preservation mechanism to adaptively sanitize participants' model updates depending on social trust degrees. \emph{iii) Distributed convergence}. A distributed two-sided matching algorithm is devised to attain an optimized disjoint partition with Nash-stable convergence. Experiments on Facebook network and MNIST/CIFAR-10 datasets validate that our SCFL can effectively enhance learning utility, improve user payoff, and enforce customizable privacy protection.
\end{abstract}


\IEEEpeerreviewmaketitle
%----------------------------------------------------------------------------------
%\textcolor{blue}{}
\section{Introduction}
With the explosive growth of smart phones, wearables, and Internet of things (IoT) devices, nearly 75\% of data is anticipated to be produced, gathered, and processed outside of clouds by 2025, particularly at distributed end-devices at the edge \cite{GartnerReport}.
Due to data privacy and ownership concerns, aggregating such vast volumes of distributed data into a central cloud for artificial intelligence (AI) model training can be both illegal and privacy risky \cite{10221755,10403801}. Federated learning (FL) offers a promising privacy-preserving AI paradigm that adheres to the principle of bringing code to data, instead of the opposite direction \cite{9928220,10292582,9159929}.
In FL, individual devices periodically train AI sub-models (e.g., gradients) using local data and send to the aggregation server (e.g., the cloud), which synthesizes a global AI model for next-round training \cite{9478223}. As users only share the learned model parameters instead of the original raw data, the privacy concerns can be significantly resolved under FL.

However, in such an open and untrusted FL environment, users' privacy can still be divulged from their trained sub-models (e.g., gradients) by sophisticated adversaries and the untrusted server, via attacks such as membership inference \cite{8835245} and model inversion \cite{Fredrikson2015CCS}. For example, experiments in \cite{NEURIPS2019} validate that clients' private training data can be stolen from the publicly shared gradients in vision and language tasks.
Existing countermeasures mainly rely on the local differential privacy (LDP) techniques \cite{9546481,9159929,9820771} due to the strict theoretic guarantees and low computation overhead, in which individuals independently sanitize their sub-models by adding random LDP perturbations, as shown in Fig.~1(a). In LDP-based approaches, the larger injected LDP noise enforces stronger privacy provisions but also entails more severe performance degradation, which eventually deteriorates individual payoff. Thereby, LDP-based FL approaches usually necessitate a tradeoff between privacy and utility.
Current efforts mainly focus on designing optimized FL approaches \cite{9252950,9562556,9740410,9609994,9261995} to strive for such a balance, while ignoring the inner and lasting connections among users, such as social relationships.

\begin{figure}[!t]\setlength{\abovecaptionskip}{-0.0cm}
\centering
  \includegraphics[width=9.4cm,height=4.45cm]{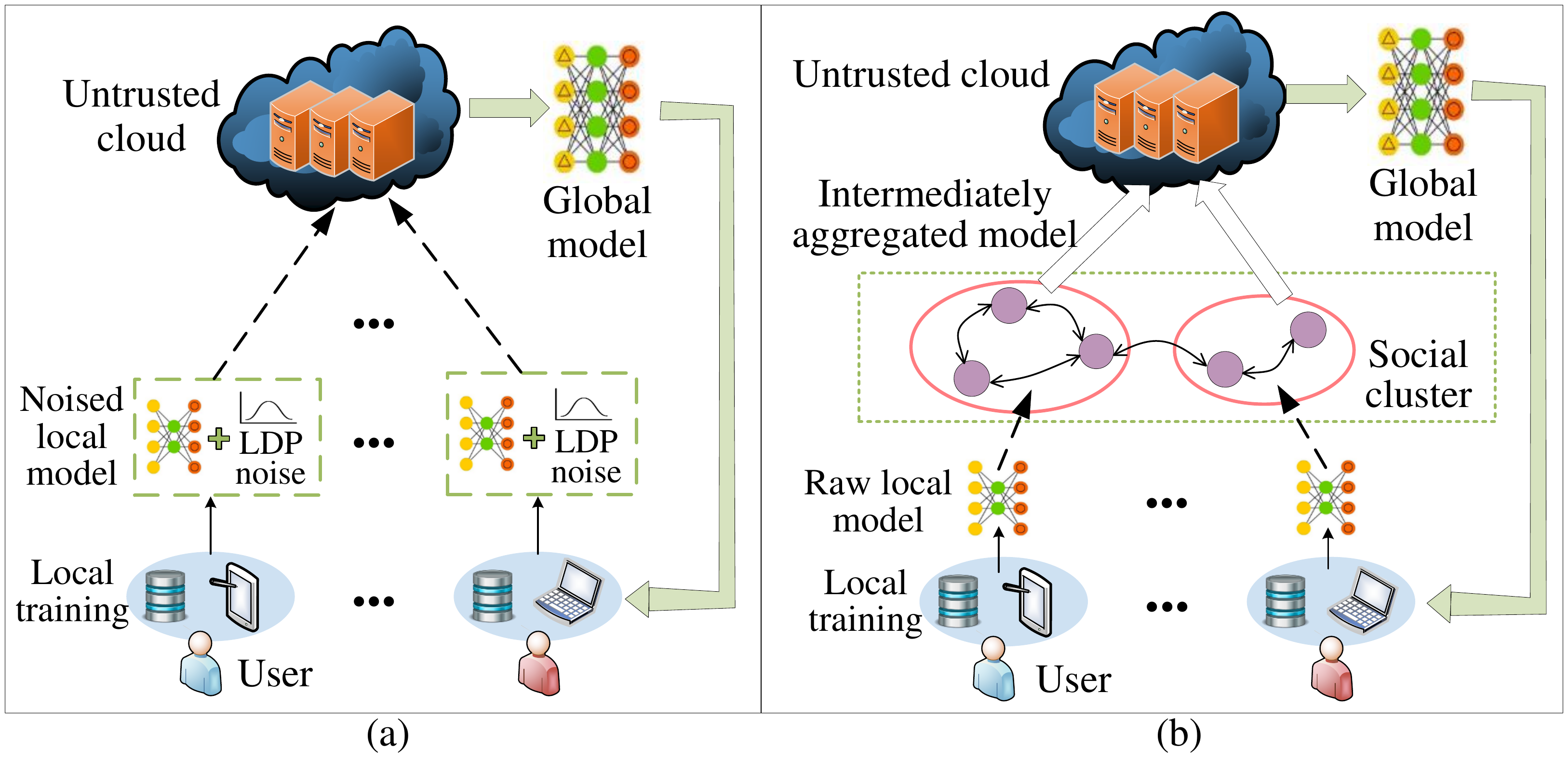}%width=9.4cm,height=4.45cm
  \caption{Illustration of (a) conventional LDP-based FL under an untrusted cloud; (b) the social-aware clustered federated learning (SCFL).}\label{fig:model}\vspace{-2mm}
\end{figure}

With the great success of social networks, social ties have been widely established among mobile users.
For instance, 2.93 billion social users monthly interacted via Facebook in the first quarter of 2022, with an average of roughly 200 friends per user \cite{10255769}.
Benefiting from large-scale social networks, individuals can easily invite their trusted and familiar social friends as cooperative learners and form socially clustered federations (or clusters), as shown in Fig.~1(b).
Within each social cluster, members can aggregate their trained sub-models into a combined one before uploading to the cloud.
Essentially, as individuals inside each social cluster are mutually trusted, they no longer need to apply LDP perturbations to the trained sub-models, which considerably enhances model utility (e.g., model accuracy) and individual payoff.
Meanwhile, both external adversaries and the curious cloud cannot deduce the gradient information of each participating user from the intra-cluster intermediate model aggregations, thus well protecting user privacy.
As such, the \underline{s}ocial-aware \underline{c}lustered \underline{f}ederated \underline{l}earning (SCFL) paradigm emerges as a promising strategy to enhance model utility while enforcing privacy protection in FL with low cost and high feasibility.

To practically deploy SCFL services, a series of fundamental challenges still need to be resolved.
1) As users are generally selfish and profit-seeking, they can determine whether or not to join a social cluster, as well as which social cluster to join, depending on the payoffs and social ties. Thereby, \emph{it remains a concern to distributively form a stable and optimized social cluster structure}.
2) Heterogeneous users typically have distinct quality, quantity, and non-IID degree of training data samples in undertaking different FL tasks, resulting in distinct model quality and contributions \cite{9478223}. Besides, selfish individuals often tend to benefit from the SCFL without contributing to the social cluster, and such free-riding behaviors may lead to poor learning outcomes~\cite{9664267}. \emph{There exists a challenge in contribution quantification and fair revenue allocation within each social cluster with free-rider defense}.
3) In clusters with low mutual social trust, learners may still need to add modest amount of LDP perturbations for privacy concerns. Due to the diversity of users' social ties and privacy preferences, \emph{how to design a flexible and differentiated perturbation mechanism to attain a tradeoff between privacy and utility is a challenging issue}.

To address these issues, this paper proposes a novel and efficient \underline{s}ocial-aware \underline{c}lustered \underline{f}ederated \underline{l}earning (SCFL) scheme with Nash-stable clustering structure, free-rider prevention, and customized privacy preservation, by using a game-theoretical approach. Specifically, we model the interactions among socially connected learners as a distributed federation game with transferable utility (FTU), and formally define the federal payoff and cost of social clusters.
Considering heterogenous training samples, data quality, and non-IID degrees of users, we then devise a fair revenue allocation mechanism for all members in each social cluster based on their quantified contributions.
Next, for users joining clusters with relatively low trust, a customizable privacy preservation mechanism is designed to meet users' privacy expectations by adaptively determining the privacy protection level depending on both trust-related factors and structural information of the social network.
Furthermore, we design an iterative two-sided matching algorithm to derive the Nash-stable social clustering structure, where each individual determines the transfer strategy to affiliate with the optimal cluster while each social cluster determines the optimal admission strategy to accept the optimal learner.

The main contributions of this work are summarized below.
\begin{itemize}
  \item \textbf{Framework}: We propose a novel hierarchical SCFL framework, which realizes low-cost, feasible, and customized FL services by exploiting users' social attributes.
  \item \textbf{Algorithms}: We formulate the optimal social cluster formation problem among individuals with social ties as a FTU game, and devise a suite of algorithms including fair revenue allocation, customizable privacy preservation, and iterative two-sided matching, which converges to Nash-stable equilibrium.
  \item \textbf{Validations}: We implement experiments on real-world datasets to validate the effectiveness of the proposed scheme. Numerical results show that our SCFL can bring higher learning utility and better individual payoff while enforcing customizable privacy protection, compared with existing representatives.
\end{itemize}

The remainder of the paper is organized as follows. Section~\ref{sec:RELATEDWORK} reviews the related works. In Section~\ref{sec:SYSTEMMODEL}, we present the system model. Section~\ref{sec:FRAMEWORK} presents the detailed construction of the proposed SCFL scheme. Performance evaluation is shown in Section~\ref{sec:SIMULATION}. Section~\ref{sec:CONSLUSION} concludes this paper and points out the future work.

\section{Related Works}\label{sec:RELATEDWORK}
%\subsection{Privacy-Utility Tradeoff in DP-Based FL}\label{subsec:relatedwork1}
In FL, many works have studied the impact of differential privacy (DP) noises on model performance, assuming that the aggregation server is semi-honest (i.e., honest-but-curious). Besides, many of them strive for a tradeoff between privacy protection and model utility under FL.
Zeng \emph{et al}. \cite{9645233} study a privacy-enhanced federated temporal difference learning mechanism by injecting DP perturbations to users' shared gradients, where both the privacy bound and the upper bound of utility loss are derived using rigorous analysis.
Wei \emph{et al}. \cite{9546481} develop an example-level DP algorithm in FL under an untrusted aggregation server, where a dynamically decaying noise-adding strategy is devised for model utility enhancement.
Shen \emph{et al}. \cite{9820771} design an optimized LDP perturbation method to reduce the impact of LDP noise in FL models via a perturbation regularizer with LDP guarantees for clients.
%Zhao \emph{et al}. \cite{9253545} design an optimal LDP mechanism under FL in vehicular networks to optimize FL performance given a large privacy budget and devise a suboptimal solution under simple situations.
Mohamed \emph{et al}. \cite{9562556} propose an optimized user sampling mechanism in DP-based wireless FL environments to balance the size of LDP noise and convergence rate.
Wei \emph{et al}. \cite{9609994} investigate a multi-agent learning approach to minimize training time and communication rounds in wireless FL settings while enforcing DP for users.
%Unlike previous works, our work introduces social effects into FL to simultaneously offer high model performance and rigorous privacy protection via social clustering and customized LDP perturbation injection with high feasibility and low cost.

%\subsection{Social-Aware Crowdsensing and FL in Wireless Networks}\label{subsec:relatedwork2}
Several works have recently been reported incorporating social effects into collaborative wireless networks such as crowdsensing and FL.
Shi \emph{et al}. \cite{9525829} exploit social influences among individuals for efficient incentive design in crowdsensing applications, with the aim to maximize the data quality of the crowdsensing platform and cut down the cost of user recruitment for data collection under information asymmetry.
By recruiting trustworthy social friends as collaborative learners, Lin \emph{et al}. \cite{9410383} design a social-driven incentive mechanism under federated edge learning to minimize the payment of FL service requesters while encouraging edge devices' resource contributions in FL.
However, existing works mainly leverage the social influences for reliable participant recruitment and cost-effective incentive design, whereas the use of social attributes and formation of social clusters among users for better privacy-utility tradeoff in DP-based FL are not taken into account.
%both the optimized social clustering structure and individual privacy protection are not taken into account.

Distinguished from previous works, our work integrates social effects into FL to simultaneously achieve high model performance and rigorous privacy protection with high feasibility and low cost, via optimized social clustering and customized LDP perturbations.

%\vspace{-1mm}
\section{System Model}\label{sec:SYSTEMMODEL}
In this section, we {first introduce the network model, and then discuss the design goals of SCFL. Table~\ref{table0} summarizes the notations used in the remaining of this paper.}

\begin{table}[!t]%\scriptsize
{
\caption{Summary of Notations}\label{table0}\centering
\vspace{-2mm}
\resizebox{1.0\linewidth}{!}{
\begin{tabular}{|c|l|}
\hline %\hline
\textbf{Notation} & \textbf{Description} \\   \hline
    $\mathcal{N}$&Set of users with social connections in a FL task. \\
    $\Phi$&Set of disjoint social clusters for users in $\mathcal{N}$. \\
    $\phi_j$&Cluster head of $\Phi_j$. \\
    $\Upsilon_{n,j}$&Social influence of user $n$ in social cluster $\Phi_j$. \\
    $\mathcal{G}$&Social graph of users in $\mathcal{N}$. \\
    $\Theta^k$&Global model at global round $k$. \\
    $\Theta_n^k$&Local model of user $n$ at global round $k$. \\
    $\alpha_{n,j}$&Social trust degree between user $n$ and $\phi_j$. \\
    $\alpha_{th}$&Predefined trust threshold. \\
    $e_{n,m}$&Direct trust between users $n$ and $m$. \\
    $\tau _{n,m}$&Indirect trust between users $n$ and $m$. \\
    $K_{n,m}^{P}$&Numbers of positive interactions between users $n$ and $m$. \\
    $K_{n,m}^{N}$&Numbers of negative interactions between users $n$ and $m$. \\
    $\nu$&Penalty factor for negative interactions. \\
    $\Gamma_b$&Exponential time decay effect. \\
    $\xi$&Time decay rate. \\
    $T_{path}$&Shortest social path connecting users $n$ and $m$. \\
    $\sigma_{n,j}$&Scale of DP noise of user $n$ in social cluster $\Phi_j$. \\
    $\sigma_{\max}$&Maximum affordable noise scale. \\
    $\epsilon_{n,j}$&Privacy budget of user $n$ in social cluster $\Phi_j$. \\
    $\delta$&Small failure possibility in DP. \\
    $\gamma$&Concentration factor of Dirichlet distribution. \\
    $q_n$&Quality of local model of user $n$.  \\
    ${\mathscr{L}}_n$&Loss value of user $n$'s trained model. \\
    $\mathcal{R}(\Phi_j)$&Federal revenue of social cluster $\Phi_j$. \\
$\mathcal{C}(\Phi_j)$&Federal cost of social cluster $\Phi_j$. \\
$\mathcal{V}(\Phi_j)$&Federal utility of social cluster $\Phi_j$. \\
    $\psi_n(\Phi_j)$&Individual payoff of user $n$ in social cluster $\Phi_j$. \\
$\varsigma$&Additional reward assigned to each cluster head. \\
$\varpi_{n,j}$&Weight of member $n$ in social cluster $\Phi_j$. \\
$\mathcal{A}$&Action space of user in social cluster formation. \\
$\rho_n(.)$&Preference function for user $n$. \\
$\mathcal{H}_n$&Historical clusters that have rejected user $n$'s transfer request. \\
    $\mathcal{C}_{n,t}^{\mathrm{cluster}}$&Transferable cluster set of user $n$ at iteration $t$. \\
    $\mathcal{C}_{\mathcal{S},t}^{\mathrm{user}}$&Set of candidate users that tend to join social cluster $\mathcal{S}$. \\
\hline %\hline
\end{tabular} } }
\end{table}

\begin{figure}[!t]\setlength{\abovecaptionskip}{-0.0cm}%\vspace{-1mm}
\centering
  \includegraphics[width=9.cm]{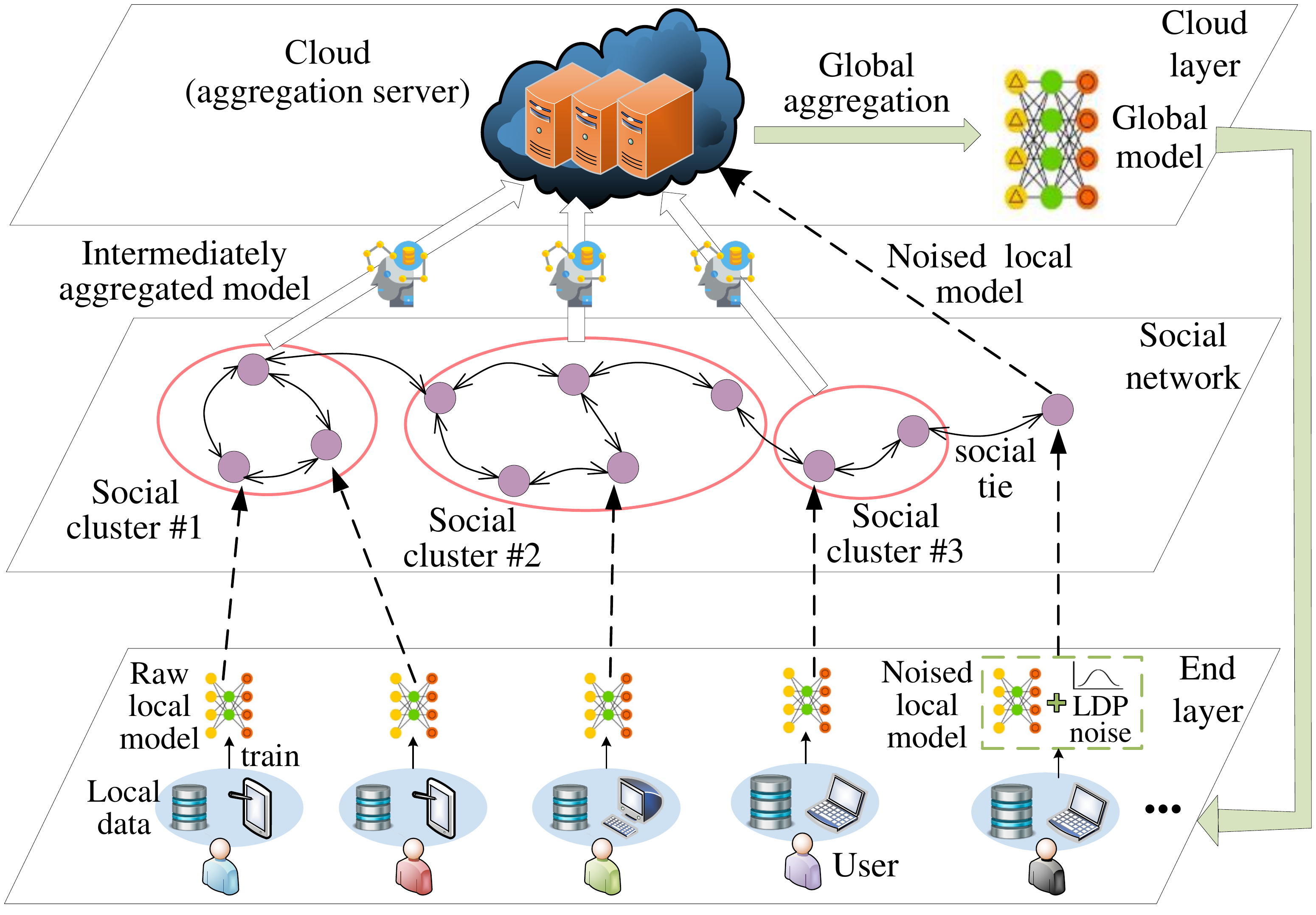}%,height=6.15cm; width=8.6cm,height=5.9cm
  \caption{Illustration of the social-aware clustered federated learning (SCFL).}\label{fig:model}\vspace{-3mm}
\end{figure}

\vspace{-2mm}
\subsection{Network Model}\label{subsec:model1}
As depicted in Fig.~\ref{fig:model}, our SCFL consists of three layers: the cloud layer, the social layer, and the end layer.

\emph{Cloud layer}. The public cloud (e.g., Azure, AWS, and Google cloud) serves as the global aggregation server in FL, which is assumed to be \emph{semi-trusted} (i.e., \emph{honest-but-curious}) \cite{9546481,9820771}. Namely, the cloud will honestly perform the global model aggregation in each communication round $k$ but is curious about the privacy in user's local model updates. The cloud platform hosts various FL tasks (e.g., image and sentiment classification) to be accomplished.
%A set of heterogeneous FL tasks (e.g., emoji prediction and sentiment analysis) to be completed are published on the cloud platform, denoted by $\mathcal{I} = \{1,\cdots,i,\cdots,I\}$.

\emph{End layer}. The end layer is composed of a set of users interested in participating in FL services, denoted by $\mathcal{N} = \{1,\cdots,n,\cdots,N\}$. %Let $\mathcal{N} \subseteq \mathcal{N}$ be the set of users who are involved in a common FL task $i\in \mathcal{I}$, where ${\left| \mathcal{N} \right|} = {N}$.
For each FL task, each user $n\in\mathcal{N}$ uses the owned smart device (e.g., smart phones, wearables, and smart vehicles) to jointly train a globally shared AI model using their local private data under the FL paradigm, coordinated by the cloud.
%For FL task, each user $n\in\mathcal{N}$

\emph{Social layer}. Generally, users are featured with social attributes (e.g., friends, relatives, and classmates) and are interconnected via the social network. In SCFL, users involved in a common task can dynamically form socially clustered disjoint federations, and the set of which is denoted as $\Phi = \{\Phi_1,\cdots,\Phi_j,\cdots,\Phi_J\}$. Within each social cluster $\Phi_j \in \Phi$, mutually trusted peers can directly send their raw local model updates instead of the noised version to the \emph{cluster head}\footnotemark[1], denoted by $\phi_j$, who then produces an intermediately aggregated model. \footnotetext[1]{A trusted processor \cite{9468910} such as Intel SGX and ARM TrustedZone can also act as the cluster head for intra-cluster model aggregation.}Here, the member with the highest centrality degree (i.e., social influence $\Upsilon_{n,j}$) in the cluster is selected as the cluster head (details are shown in Sect.~\ref{subsec:model2}), which is assumed to be \emph{socially trusted}\footnotemark[2] inside the social cluster. \footnotetext[2]{In a social cluster, its security level (e.g., risk of user's privacy leakage) in social-layer model aggregation is associated with the mutual social trust between the user and the cluster head. As the social trust degree may vary over time and across users, the security level of the social cluster can be temporally dynamic and heterogenous for different users.}After aggregating the local models, the cluster head forwards the combined model to the cloud. As such, both the global model's utility (e.g., model performance) and users' payoffs can be improved. Besides, since neither the curious cloud nor external adversaries can infer any member's raw model update from the intermediate model aggregation, the privacy of learners in each social cluster can be well-protected (details are shown in Sect.~\ref{subsec:scheme4}).

{In a typical cross-device FL setting, clients only communicate with the central server. Compared with conventional cross-device FL scenarios, our SCFL framework incorporates the social relationships between users and allows socially trusted users to form stable social clusters via social platforms (e.g., Facebook) to improve model utility, which can be applied to general cross-device FL applications.}

{\emph{Illustrating example}. We take Facebook as an example of the social platform. A FL task publisher announces his/her FL task along with the Facebook group ID, learning model structure, IP address of the cloud aggregation server and etc, on a machine learning community (e.g., Kaggle). Participants can invite their Facebook friends to join the FL process. All participants form a grand Facebook group, within which they can communicate. Then, mutually trusted clients can create multiple disjoint Facebook subgroups. They can freely move between these subgroups or choose to act independently based on individual benefits, until forming a stable partition for all users. Within each subgroup, mutually trusted peers directly share their raw local model updates instead of the noised version with the cluster head, who serves as the manager of the corresponding Facebook subgroup. Each cluster head then forwards the intra-subgroup combined model to the aggregation server who produces a global model via inter-subgroup aggregation. Finally, the global model is distributed back to the grand Facebook group for next-round training.}

%\vspace{-2mm}
\subsection{Design Goals}\label{subsec:model6}
The target of our SCFL is to simultaneously attain the following design goals.

\textbf{1) Dynamically optimized social cluster structure}. Considering users' diverse social ties and dynamic competition and cooperation, SCFL aims to form an optimized social cluster structure to maximize the payoffs of profit-driven users.

\textbf{2) Fair revenue allocation with free-rider resistance}. Considering diverse user characteristics (e.g., training samples, data quality, and data distribution) and potential free-riding behaviors in FL \cite{9664267}, SCFL should resist free-riders and ensure a fair division of cluster revenue inside each social cluster.

\textbf{3) Customized privacy preservation}. For the social clusters with low mutual trust, SCFL needs to enhance the privacy protection of participants by adding customized LDP noises to their local model updates.

%\vspace{-2mm}
\section{SCFL: Social-Aware Clustered Federated Learning Scheme}\label{sec:FRAMEWORK}
{In this section, we first present the design overview (Sect.~IV-A), and then the social trust model (Sect.~IV-B) and the federation game (Sects. IV-C), followed by the game analysis (Sect. IV-D) and algorithm design (Sect. IV-E).}

%\vspace{-3mm}
\subsection{Design Overview}\label{subsec:scheme0}
{The overall objective of SCFL is to learn a global model for all participating clients.}
The workflow of SCFL consists of five successive phases: (i) social trust evaluation, (ii) social cluster formation, (iii) user-side local model training {by all clients}, (iv) intra-cluster intermediate model aggregation {by the cluster head}, and (v) inter-cluster global model aggregation by the cloud.

\underline{Phase 1: Social trust evaluation}. Each user $n\in \mathcal{N}$ evaluates the social trust degrees of existing/potential social friends in the social graph $\mathcal{G}$ via (\ref{eq:globalTrust}). {Details are shown in Sect. IV-B.}

\underline{Phase 2: Social cluster formation}. A group of socially connected users (i.e., $\mathcal{N}$) self-organize into disjoint social clusters (i.e., $\Phi$) depending on the payoffs and social trusts. Particularly, each user $n$ determines which social cluster to join or work alone. For some clusters with low mutual social trust, the participants will add modest LDP noises for privacy concerns. Besides, the members of each cluster independently decide whether to accept the newcomers. {Details are shown in Sects. IV-C$\sim$IV-E.}

\underline{Phase 3: Local model training}. Each user $n$ trains the received global AI model $\Theta^{k-1}$ in previous round $k-1$ using local private data $\mathcal{D}_n$ via stochastic gradient descent (SGD) and produces the local AI sub-model $\Theta_n^k$ in current round:
\begin{align}\label{eq:localTrain}
\Theta_n^k \leftarrow \Theta^{k-1} - \eta \nabla{\mathscr{L}_n}\left( \Theta^{k-1} \right),
\end{align}
where $\eta$ is the learning rate and ${\mathscr{L}_n}$ is the loss function on user $n$'s local data samples.
If user $n$ joins a cluster $\Phi_j$ with high social trust (i.e., $\alpha_{n,j}\ge \alpha_{th}$), the raw sub-model $\Theta_n^k$ is directly sent to the cluster head $\phi_j$. If user $n$ joins a cluster with relatively low trust (i.e., $0<\alpha_{n,j}<\alpha_{th}$), a modestly noised sub-model $\widetilde{\Theta}_n^k$ is sent to $\phi_j$. Here, $\alpha_{n,j}$ is the trust degree between user $n$ and $\phi_j$, and $\alpha_{th}\in (0,1)$ is a predefined public trust threshold, whose value depends on specific FL tasks. Otherwise, user $n$ works alone and uploads the noised model $\widehat{{\Theta}}_n^k$ injected with the uniform and relatively large LDP noise to the cloud (details are shown in Sect.~\ref{subsubsec:customized}).

\underline{Phase 4: Intra-cluster model aggregation}. Each social cluster head $\phi_j$ aggregates the local model updates of all the members\footnotemark[3] in the social cluster and uploads the intermediate aggregation outcome to the cloud, i.e.,
\begin{numcases}{}
\overline{\Theta}_j^k = \sum\nolimits_{n\in {\mathcal{N}_{j}^1}}{q_{n,j}\Theta_n^k} + \sum\nolimits_{n\in {\mathcal{N}_{j}^2}}{q_{n,j}\widetilde{\Theta}_n^k},\label{eq:SocialAggre1} \hfill \\[-0.1pt]
{Q_{j}} = \sum\nolimits_{n\in {\mathcal{N}_{j}^1}\bigcup {\mathcal{N}_{j}^2}}{q_{n,j}}. \label{eq:SocialAggre2}
\end{numcases}
In (\ref{eq:SocialAggre1}) and (\ref{eq:SocialAggre2}), ${\mathcal{N}_{j}^1}$\,and $\!{\mathcal{N}_{j}^2}\!$ are the sets of users that send raw sub-models and noised sub-models to cluster $\Phi_j$, respectively. $q_{n,j}$ is the quality of user $n$'s local model, which is evaluated in Sect.~\ref{subsubsec:quality}. {When detailed curve-fitting parameters are unavailable, we adopt an iterative approach in \cite{9488743} (Sect. IV-A-1) to estimate the quality of client's local model.} \footnotetext[3]{To resist Byzantine attacks (e.g., model poisoning) of participants, existing Byzantine-resilient aggregation mechanisms \cite{9757841,9887909,9468910,9911773} in different FL settings can be further applied, which is out of scope of this paper.}

\underline{Phase 5: Inter-cluster global model aggregation}. The cloud synthesizes the intermediate aggregations from various social clusters into the current global model $\Theta^{k}$ weighted by model utilities, i.e.,
\begin{align}\label{eq:globalaggregation}
\Theta^{k} \leftarrow \frac{1}{\sum\nolimits_{\Phi_j \in \Phi}{Q_{j}}}{\sum\nolimits_{\Phi_j \in \Phi} \overline{\Theta}_j^k}.
\end{align}

Until the global round $k$ attains its maximum value or the global model obtains the predefined accuracy, the above learning process in phases 3-5 finishes.

\subsection{Social Trust Evaluation}\label{subsec:model2}
Let $\mathcal{G} = \left<\mathcal{N},\mathcal{E},\mathcal{T}\right>$ denote the social graph among users in the set $\mathcal{N}$. Here, $\mathcal{E}=\{e_{n,m}| \forall n,m \in \mathcal{N}, n\ne m\}$ is the set of edges between users, and $e_{n,m} \in [0,1]$ denotes the social relationship or social closeness between two users $n$ and $m$ ($n\ne m$). %with $e_{n,m}=e_{m,n}$.
Particularly, $e_{n,m}=1$ means that two users have the strongest social tie, and $e_{n,m}=0$ implies that they are strangers. Let $\alpha_{n,m}\in [0,1]$ denote the social trust degree between two users $n$ and $m$, and the set of which is denoted as $\mathcal{T}=\{\alpha_{n,m}| \forall n,m \in \mathcal{N}, n\ne m\}$.
The social trust degree $\alpha_{n,m}$ is evaluated based on the direct social closeness and indirect topological relationships{\cite{6226368,8678450}}.

The direct trust $e_{n,m}$ is oriented from the direct experience in historical interactions (e.g., sharing microblogs, photos, videos, and engaging in social gaming){\footnotemark[4]}, which is affected by the interaction experience and interaction occurrence time. {According to \cite{8358773}, we have}
{\begin{align}\label{eq:directTrust}
e_{n,m}\!=\!\max \bigg\{\frac{\sum_{b=1}^{K_{n,m}^{P}}{\Gamma_b}-\nu \sum_{b=1}^{K_{n,m}^{N}}{\Gamma_b}}{ K_{n,m}^{P}+K_{n,m}^{N} }, 0 \bigg \},
\end{align}}
where $K_{n,m}^{P}$ and $K_{n,m}^{N}$ are the total numbers of positive and negative interactions{\footnotemark[5]} between user $n$ and user $m$, respectively. $\nu>0$ is a penalty factor.
$\Gamma_b=\exp \left( -\xi \left( t-t_b \right) \right)$ describes the exponential time decay effect {as latest interaction can be more important than older ones}, where $t_b$ is the occurrence time of $b$-th interaction and $\xi>0$ is the decay rate.\footnotetext[4]{{It is assumed that for all users engaging in a common FL task, their social trust degrees do not update until a Nash-stable partition in Alg.~1 is formed.}}\footnotetext[5]{{The positive and negative interactions are determined based on user's subjective feelings {(e.g., giving a subjective rating to his/her peer)} during each {online or offline} interaction, which may not be symmetric.}}

As direct interactions between users are often inadequate, combining indirect topological relationships {(i.e., friend-of-friend relationships)} in the social graph is necessary for comprehensive trust evaluation. Let $T_{path}$ denote the shortest path connecting user $n$ and user $m$ in $\mathcal{G}$, {which excludes the direct link.} $|T_{path}|$ is called the social distance{\footnotemark[6]}.\footnotetext[6]{{We set $|T_{path}|=2$ to obtain the social recommendation only from his/her friends to preserve user privacy to a large extent. If multiple paths share the same social distance, the indirect trust score is computed by averaging the aggregated recommendations on these pathes. {Notably, the computing of indirect trust between two users involves all their common friends in $\mathcal{G}$, regardless of their participation status in the FL task.}}}
The indirect trust can be computed as the aggregated recommendations from his/her friends in $T_{path}${\cite{8762069}}, i.e.,
\begin{align}\label{eq:indirectTrust}
\tau _{n,m}=\prod_{l,k\in T_{path},l\mapsto k}{e_{l,k}}.
\end{align}
$l\mapsto k$ means that users $l$ and $k$ are adjacent in the path $T_{path}$.
{Notably, the multiplication of trust values is adopted instead of computing the average (as done in \cite{6226368}) to reflect the impact of a low trust value on the global aggregation outcome.}

Thereby, the global social trust degree can be attained as:
\begin{align}\label{eq:globalTrust}
\alpha _{n,m}=\omega e_{n,m} + (1 - \omega)\tau _{n,m},
\end{align}
where $\omega\in [0,1]$ is the weight factor. Notably, $\alpha _{n,m}\in [0,1]$. Besides, as the social connections between users are temporally evolutionary, the social trust $\alpha _{n,m}$ between users is dynamically evaluated.
Denote $\Upsilon_{n,j}$ as the centrality degree (or social influence) of user $n$ in social cluster $\Phi_j$, i.e., the number of neighbors that user $n$ has in cluster $\Phi_j$. Here, $\Upsilon_{n,j} = \sum_{l \in \Phi_j}{f_{n,l}}$, where ${f_{n,l}}=\{0,1\}$ and ${f_{n,l}}=1$ if $e_{n,l}>0$. Otherwise, ${f_{n,l}}=0$.

\subsection{Federation Game Formulation}\label{subsec:scheme3}
In federation game, the mutual communication capability among all players is a basic assumption, which the social network in our work can enable.
The social cluster formation process among social individuals is formulated as a federation game with transferable utility (FTU), where socially connected learners can self-organize into disjoint social clusters for maximized individual payoffs.
\\
\emph{\textbf{Definition 1} (FTU game):}
For every FL task, a FTU game is formally defined by a 4-tuple $(\mathcal{N}, \Phi,\mathcal{V},\mathcal{A})$, which includes the following key components.
\begin{itemize}
  \item \emph{Players:} The game players are a set of social users involved in FL task (i.e., $\mathcal{N}$).
  \item \emph{Federation structure:} A partition structure, denoted as $\Phi = \{\Phi_1,\cdots,\Phi_J\}$, divides the player set $\mathcal{N}$ into mutually disjoint clusters such that ${\Phi_j}\cap{\Phi_{j'}}=\emptyset$, $\forall j \ne j',\forall i \in \mathcal{I}$, and $\cup_{j=1}^{J}{\Phi_j}=\mathcal{N}$.
  \item \emph{Payoff:} The federal payoff of each social cluster $\mathcal{S} \subseteq \mathcal{N}$, denoted as $\mathcal{V}(\mathcal{S})$, can be arbitrarily apportioned among all the members within $\mathcal{S} \in \Phi$. The individual payoff of each player $n \in \mathcal{N}$ that joins in a cluster $\mathcal{S} \in \Phi$ is denoted as $\psi_n(\mathcal{S})$.
  \item \emph{Strategy:} The action space of each player is denoted as $\mathcal{A}$. Each player can determine either to act alone by applying the \emph{solo training} strategy or join a social cluster to jointly produce a intra-cluster aggregated model using the \emph{clustered training} strategy.
\end{itemize}

Next, we define group rationality and individual rationality. Based on them, the core of the FTU game is defined.
\\
\emph{\textbf{Definition 2}:} A payoff vector $\bm{\psi} = \left\{\psi_n \right\}_{n=1}^{N}$ is said to be \emph{group rational} if $\sum_{n=1}^{N}{\psi_n} = \mathcal{V}(\mathcal{N})$. Besides, $\bm{\psi}$ is said to be \emph{individual rational} if $\psi_n \ge \mathcal{V}(\{n\}), \forall n \in \mathcal{N}$, i.e., the payoff of any user in the FTU game is no less than what they would receive from acting alone.
\\
\emph{\textbf{Definition 3}:} The \emph{core} of the FTU game is a set of stable payoff vectors satisfying both group rationality and individual rationality, i.e.,
\begin{align}\label{eq:core}
\mathcal{C}\!=\! \left \{\bm{\psi} | \sum_{n=1}^{N}{\psi_n} \!=\! \mathcal{V}(\mathcal{N}) \,\& \sum_{n\in \mathcal{S}}{\psi_n} \!\ge\! \mathcal{V}(\mathcal{S}),\forall \mathcal{S}\subseteq \mathcal{N} \right \}\!.\!
\end{align}
In (\ref{eq:core}), $\sum_{n\in \mathcal{S}}{\psi_n} \!\ge\! \mathcal{V}(\mathcal{S})$ means that players have no incentives to form another cluster $\mathcal{S}$ and reject the proposed $\bm{\psi}$.
A non-empty core implies that participants are incentivized to form the grand federation (i.e., $\{{N}\}$).

\vspace{-2mm}
\subsection{Federal and Individual Payoff Analysis}\label{subsec:scheme4}
\subsubsection{Customized Local Perturbation}\label{subsubsec:customized}
In the case that a user joins a specific cluster with relatively low mutual social trust, the user may still need to add a modest amount of LDP perturbations for privacy concerns. In most previous works \cite{9546481,9928220,9645233}, it is supposed that all users are subject to a uniform level of privacy protection, which rules out users' personalized privacy preferences. Here, we develop a trust-oriented customized local perturbation mechanism to satisfy individual privacy expectations in practical scenarios. Particularly, the Gaussian mechanism is adopted by adding artificial noise following the Gaussian distribution ${\mathbb{G}}(0,\sigma^2 S^2)$. The variance $\sigma$ controls the scale of noise. Based on the moments accountant method \cite{Abadi2016Deep}, to preserve $(\epsilon,\delta)$-LDP, the noise scale should satisfy \cite{9820771}:
\begin{align}\label{eq:noisescale}
\sigma \geq \frac{\sqrt{2\log(1.25/\delta)}}{\epsilon}.
\end{align}
In (\ref{eq:noisescale}), $\epsilon>0$ is the privacy budget, and a smaller $\epsilon$ enforces stronger privacy protection. $\delta$ is a small failure possibility (we set $\delta=10^{-6}$).
Besides, $S=\max_{\mathcal{D},\mathcal{D}'}||f\left( \mathcal{D} \right) -f\left( \mathcal{D}' \right) ||_2$ is the L2-sensitivity of query function $f$ on two neighboring datasets $\mathcal{D}$ and $\mathcal{D}'$.
Let $\lambda_s$ be the sampling rate of user's local data samples. In the following theorem, we show the privacy amplification property of DP.
\\
\emph{\textbf{Theorem 1}{\cite{Balle2018Privacy}}:}
According to the privacy amplification property, the Gaussian mechanism with sub-sampling ensures $(\epsilon',\lambda_s \delta)$-LDP and guarantees stronger privacy preservation, where
\begin{align}\label{eq:epsilon-LDP}
\epsilon' = \log(1 + \lambda_s(\exp(\epsilon) - 1)).
\end{align}
%\begin{IEEEproof}
%The detailed proof can refer to \cite{Balle2018Privacy}.
%\end{IEEEproof}

\emph{Remark:} Theorem~1 shows that it provisions stronger privacy preservation by applying the DP mechanism on a random subset of participant's local data samples than on the entire dataset. Moreover, Theorem~1 indicates that the added differentiated Gaussian noises strictly enforce LDP and can preserve participants' privacy in social-layer model aggregation process.

Whenever a user $n$ tends to join a cluster ${\Phi_j}$ with $\alpha_{n,j}\in (0,\alpha_{th})$, our mechanism returns a sanitized AI sub-model $\widetilde{\Theta}_n^k$ in which the chosen privacy level depends on his/her social trust degree $\alpha_{n,j}$ with the corresponding cluster head $\phi_j$. Specifically, the customizable privacy budget level in LDP can be linearly mapped based on corresponding trust degree, i.e.,
\begin{align}\label{eq:CustomizedLDP}
\epsilon_{n,j}= \theta_1 \cdot \frac{\alpha _{n,j}}{\alpha _{n,j} + \theta_2},
\end{align}
where $\theta_1$ and $\theta_2$ are positive adjustable coefficients.

\subsubsection{Local Update Quality Evaluation}\label{subsubsec:quality}
In SCFL, users usually have distinct data sizes and distributions, as well as the injected Gaussian noise scales on local model updates when joining clusters with relatively low trust. Typically, the lower scale $\sigma_{n,j}$ of injected Gaussian noise, the better performance of the trained model. Besides, as validated in \cite{9155494}, non-IID data can cause performance deterioration in FL compared with IID data. {We consider the \emph{label- and quantity-skewed non-IID setting} \cite{lin-etal-2022-fednlp} and focus on \emph{classification tasks} under the FL paradigm. In the literature, the Dirichlet distribution has been widely adopted for dataset partition with both quantity and label distribution shifts under the non-IID environment for FL classification tasks, such as \cite{Hsu2019MeasuringTE} for image classification tasks and \cite{lin-etal-2022-fednlp} for text classification tasks. Hence,} we employ the Dirichlet distribution to characterize the heterogeneity of data size and data distribution among FL participants. Consider a classification task with $Y$ classes, where training examples of each client is drawn from a Dirichlet distribution parameterized by a vector $\bm{a} \sim \mathrm{Dir}(\bm{\gamma})$ with the following probability density function (PDF):
\begin{align}
p(\bm{a}|\bm{\gamma}) = \frac{1}{\mathcal{B}(\bm{\gamma})}\prod_{y=1}^Y a_y^{\gamma_y -1},~\gamma_y>0,\ \sum_{y=1}^{Y}{a_y}=1,
\end{align}
where the multivariate beta function $\mathcal{B}(\bm{\gamma}) \triangleq \frac{\prod_{y=1}^Y\Gamma(\gamma_y)}{\Gamma(\sum_{y=1}^Y \gamma_y)}$ is the normalization constant. $\gamma_y>0, \forall y\in [1,Y]$ is a concentration factor controlling the identicalness among participants. If $\gamma_y \rightarrow \infty ,\forall y$, all users have identical distributions. If $\gamma_y \rightarrow 0,\forall y$, each user only holds one class of samples at random.
For simplicity, we set $\gamma_y=\gamma,\forall y${\footnotemark[7]}.

Real-world experimental results on the MNIST dataset in Sect.~\ref{subsec:evalution2} show that the {test loss value ${\mathscr{L}}_n$ of the distributively trained model\footnotemark[8] at the end of training (i.e., when the communication rounds reach the maximum value)} can be well-suited by the 3D sigmoid curve with respect to user $n$'s noise scale $\sigma_{n,j}$ and non-IID degree ${\gamma}$:
\begin{align}\label{eq:lossfitting}
{\mathscr{L}}_n = {\mathscr{L}}(\sigma_{n,j},\gamma)= \frac{\mu_1 \exp(-\mu_2 \cdot \gamma)}{\mu_3 + \exp(-\mu_4 \cdot\sigma_{n,j})} + \mu_5,
\end{align}
where $\mu_1,\cdots,\mu_5$ are positive curve fitting parameters{\footnotemark[9]}. {The loss in Eq (13) is an empirical fit to the actual loss.} The numerator part $\mu_1 \exp(-\mu_2 \cdot \gamma)$ reflects the diminishing marginal loss value when the concentration factor $\gamma$ increases. The denominator part ${\mu_3 + \exp(-\mu_4 \cdot\sigma_{n,j})}$ captures that the increasing noise scale $\sigma_{n,j}$ results in a performance degradation.
Notably, curve fitting is a typical approach to determine the AI model quality, and a similar manner has been applied in works \cite{9094030,8486241}. In the experiments in Sect.~\ref{subsec:evalution2}, the loss function in (\ref{eq:lossfitting}) fits well when $\sigma_{n,j}$ falls in $[0,\sigma_{\max}]$, where $\sigma_{\max}$ is the maximum tolerable noise scale to guarantee model availability in practical FL services. It is because that oversized noise can completely distort model parameters and lower model accuracy to the level of random inference.\footnotetext[7]{{In practice, the value of $\gamma$ can be estimated by the aggregation server before performing the FL task. For example, a questionnaire can be sent to all participating users, which collects their data distribution (including the label classes of local data and the corresponding data amount for each class). After computing the frequency of the class for each user's dataset, the approximate value of $\gamma$ in Dirichlet distribution can be estimated via parameter estimation methods such as maximum likelihood estimation \cite{minka2000estimating,StatProofBook}.}}\footnotetext[8]{{When clients adopt different DP noise scales under FL, it becomes very challenging to directly obtain the theoretical relationship between the local model quality of client $n$ and its noise scale. As an alternative, we use the quality of the global model, where all clients adopt the same DP noise scale $\sigma_{n,j}$, to represent the local model quality of client $n$ with noise scale $\sigma_{n,j}$.}}\footnotetext[9]{{As the curve-fitting parameters $\mu_i$ are fixed and uniform for all users involved in a common FL task, the stable social cluster structure produced by the federated game algorithm in Alg.~1 does not depend on the detailed values of $\mu_i$ but the form of curve-fitting function. As the specific form of curve-fitting function ${\mathscr{L}}_n$ can be a priori for a given FL task, we can leverage the historical knowledge for various FL tasks in the public FL market\cite{9094030}. Specifically, 1) if the target FL model exists in the historical FL tasks, we directly apply the corresponding form of curve-fitting function; 2) otherwise, we can select a historical FL task that closely resembles the target task, and employ the corresponding curve-fitting form.}}

Typically, the lower the loss value, the better the model performance. Moreover, the faster the drop rate of the loss, the faster the ascent rate of the model utility (e.g., model accuracy). Hence, the model utility function ${q}({\mathscr{L}}_n)$ should meet ${q}({\mathscr{L}}_n) >0$ and $\frac{\mathrm{d} {q}({\mathscr{L}}_n)}{\mathrm{d}{\mathscr{L}}_n}<0$. Via curve fitting approaches, the quality function with respect to the loss value is formulated in the linear form to meet the above requirements, i.e.,
\begin{align}\label{modelquality}
q_n={q}({\mathscr{L}}_n) = -\kappa_1 \cdot {\mathscr{L}}_n + \kappa_2,
\end{align}
where $\kappa_1>0$ is a positive adjustment factor. The factor $\kappa_2$ captures the maximum model utility when the loss ${\mathscr{L}}_n\rightarrow 0$.

\subsubsection{Federal Payoff Function}\label{subsubsec:federalpayoff}
The total revenue of social cluster $\Phi_j$ is related to the overall utility of immediate model aggregations. For simplicity, the federal revenue is computed based on the sum of model utilities of all its members \cite{7093125}:
\begin{align}\label{eq:5-1}
\mathcal{R}(\Phi_j)= \lambda_p \sum\nolimits_{n\in \Phi_j}{q_{n}},
\end{align}
where $\lambda_p$ is the task publisher's unit payment per model quality. ${q_{n}}$ is the quality of local model (QoLM) of user $n$ in cluster $\Phi_j$ based on (\ref{eq:lossfitting}) and (\ref{modelquality}). Besides, individuals can apply the solo training strategy by forming a singleton, namely, $\Phi_j=\{n\}$. In this case, the federal revenue is computed as $\mathcal{R}(\{n\})= \lambda_p {\widehat{q}_n}$, where ${\widehat{q}_n}$ is computed via $\sigma_{n,j}=\sigma_{\max}$.

Users within each social cluster need to frequently communicate with the cluster head for intra-cluster model aggregation.
Based on \cite{5678781,6550882}, the federal cost $\mathcal{C}(\Phi_j)$ can be measured by the communication overhead that is proportional to the cluster size ${|\Phi_j|}$, i.e.,
\begin{align}
\mathcal{C}(\Phi_j) =
\begin{cases}
\lambda_c{|\Phi_j|}, & \mathrm{if}\ {|\Phi_j|}> 1,\\
0, & \mathrm{if}\ {|\Phi_j|}=1,
\end{cases}
\end{align}
where $\lambda_c$ is a positive scaling coefficient.

The federal utility of social cluster $\Phi_j$ can be denoted as the total revenue minuses the cost:
\begin{align}
\mathcal{V}(\Phi_j) &=\mathcal{R}(\Phi_j) - \mathcal{C}(\Phi_j) \nonumber \\
&=\begin{cases}
\lambda_p \sum_{n\in \Phi_j}{q_n} - \lambda_c{|\Phi_j|}, & \mathrm{if}\ {|\Phi_j|}> 1,\\
\lambda_p \widehat{q}_n, & \mathrm{if}\ {|\Phi_j|}=1.
\end{cases}
\end{align}

\begin{figure}[!t]\setlength{\abovecaptionskip}{-0.0cm}
\centering
  \includegraphics[height=4.8cm]{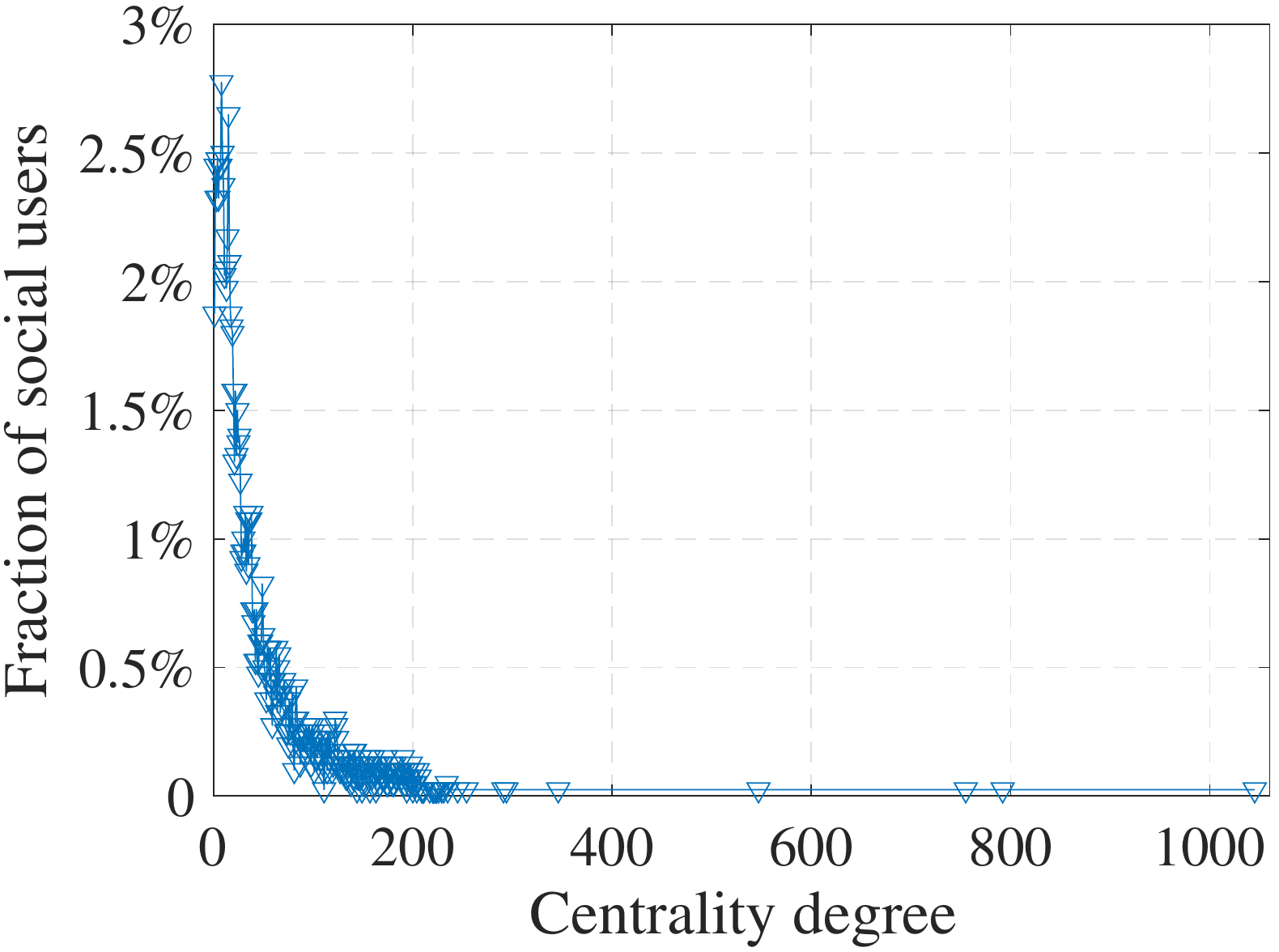}%width=5.2cm,
  \caption{Node centrality degree distribution of Facebook network \cite{FaceBookEgo}.}\label{fig:degreeDis}\vspace{-2.5mm}
\end{figure}
\subsubsection{Fair Payoff Division Within Cluster}\label{subsubsec:payoffdivision}
The proportional fairness is employed for fair payoff division within each social cluster while conserving individual rationality, in which the \emph{extra payoff} is divided into weights based on participants' non-cooperative payoffs. The individual payoff of user $n \!\in\! \Phi_j$ is given by
\begin{align}\label{eq:PF}
\psi_n(\Phi_j) = \varpi_{n,j} \left( \mathcal{V}(\Phi_j) - \sum\nolimits_{l \in \Phi_j}{\mathcal{V}(\{l\})} - \varsigma \right) + {\mathcal{V}(\{n\})},
\end{align}
where ${\mathcal{V}(\{l\})}$ and ${\mathcal{V}(\{n\})}$ denote the non-cooperative payoffs of users $l$ and $n$, respectively. $\varpi_{n,j} = \frac{q_{n}}{\sum_{l \in \Phi_j}{q_{l}}}$ is the weight of member $n$ in cluster $\Phi_j$. In (\ref{eq:PF}), users with higher QoLMs deserve more extra revenue.
Besides, as the cluster head is responsible for intra-cluster model aggregation and result uploading, it consumes more computation, storage, and communication resources than other individuals within the cluster. Thereby, an additional reward $\varsigma$ is assigned to the cluster head as an incentive.

\emph{Remark:}
When performing different FL tasks, as participants usually have different data sizes, data quality, and data distributions, they usually form different social clusters.

{Next, we show that the grand social cluster $\{{N}\}$ which contains all users in $\mathcal{N}$ will not form for a FL task.} In the FTU game, the federal cost of a cluster grows as the cluster size increases, which can be considerably high for the grand federation. Besides, it is usually impractical for all users to maintain a close social connection with the cluster head $\phi$ of the grand federation. As depicted in Fig.~\ref{fig:degreeDis}, experiments on the real-world Facebook network \cite{FaceBookEgo} with 4039 nodes and over 88K edges show that the top-4 users with the highest centrality have 1045, 792, 755, and 547 social neighbors, respectively, and similar results also apply for other social networks such as Google+ and Twitter networks. It implies that social users generally belong to different social communities. If the grand federation exists, due to the relatively low social trust with the cluster head $\phi$, most users still need to add modest LDP noise via (\ref{eq:CustomizedLDP}) in the grand federation, which eventually deteriorates model utility and individual payoff. As such, part of users tend to deviate from the grand coalition and form several disjoint clusters. Hence, the grand federation is unstable. %and Theorem 2 is proved.

{\emph{Remark:} According to Definition~3, as the grand social cluster will not form, the core of $(\mathcal{N}, \Phi,\mathcal{V},\mathcal{A})$-FTU game is empty. It} indicates that social users have no incentives to form a grand federation. In the following, we devise a distributed cluster formation algorithm to derive such stable disjoint clusters.

\subsection{Distributed Stable Social Cluster Formation}\label{subsec:scheme5}
The solution of the FTU game is to find a stable clustering structure $\Phi^*$, where the stable partition structure can be acquired via exhaustive searching \cite{9928220}. However, with more users involved, the number of possible partition iterations increases exponentially \cite{5678781}.
Alternatively, Algorithm~\ref{Algorithm1} shows an iterative two-sided matching algorithm with low complexity to distributively attain the optimal partition strategy, consisting of three steps as below.

\begin{algorithm}[t!]\begin{small}
   \caption{\small{Distributed Stable Social Cluster Formation}}\label{Algorithm1}
   \LinesNumbered
   \KwIn{$\mathcal{G}$, $\mathcal{N}$, $\mathcal{V}$, $\mathcal{A}$, $\gamma$, $\sigma_{\max}$, $\alpha_{th}$, $\omega$}
   \KwOut{$\Phi^*$, $\bm{\psi} = \left\{\psi_n \right\}_{n\in \mathcal{N}}$}
   \textbf{Initialize:} $t\!=\!0$, $\Phi\!=\!\Phi^{(0)}$, $\mathcal{C}_{n,t}^{\mathrm{cluster}}\!=\!\emptyset$, $\mathcal{C}_{\mathcal{S}}^{\mathrm{user}}\!=\! \emptyset$, $\mathcal{H}_n \!=\! \emptyset$\;
   \While{$\rho_n(\mathcal{S}') < \rho_n(\mathcal{S}), \forall n \in \mathcal{S} \subseteq\mathcal{N}, \forall \mathcal{S}'\in {\Phi}^{(t)}\cup\{\emptyset\}$}{
        \For{$n \in \mathcal{N}$}{
           Update $\mathcal{H}_n$\;
           \For{ $\Phi_j^{(t)} \in \Phi^{(t)}\setminus \mathcal{H}_n$ }{
            Compute $\alpha _{n,j}$ via (\ref{eq:globalTrust})\;
            \If{$0<\alpha_{n,j}<\alpha_{th}$}{
                Compute $\epsilon_{n,j}$ via (\ref{eq:CustomizedLDP}) and $\sigma_{n,j}$ via (\ref{eq:noisescale})\;
            \eIf{$\alpha_{n,j}=0$ or $\Phi_j^{(t)}=\{\emptyset \}$}{
                Set $\sigma_{n,j} = \sigma_{\max}$\;
                }{
                Set $\sigma_{n,j} = 0$\;
            }
            Compute $q_n$ via (\ref{eq:lossfitting}) and (\ref{modelquality})\;
        }
        Compute $\rho_n(\mathcal{S})$ via (\ref{preferencefunction}) and $\mathcal{C}_{n,t}^{\mathrm{cluster}}$ via (\ref{clusterset})\;
        Send transfer request to the cluster ${\mathcal{S}^*}\in \mathcal{C}_{n,t}^{\mathrm{cluster}}$ via the transfer rule and membership leaving rule\;
        }
      }
       \For{ ${\mathcal{S}} \in \Phi^{(t)}$ }{
            Store the users that request to transfer to it in $\mathcal{C}_{\mathcal{S},t}^{\mathrm{user}}$\;
            Accept the most preferred user $n^*\in \mathcal{C}_{\mathcal{S},t}^{\mathrm{user}}$ via the admission rule and membership joining rule while reject other candidates\;
            Do split-and-merge operation via Definition~9\;
        }
        $t = t + 1$\;
   }\end{small}
\end{algorithm}

\underline{Step 1: Partition initialization}~(line 1). For each FL task, the initial partition ${\Phi ^{(0)}}$ at $t=0$ depends on specific applications, such as the stable partition results of the previously completed FL mission. {When previous partitions are not available, the initial social clustering structure is set as $\Phi ^{\left( 0 \right)}=\left\{ 1,2,\cdot \cdot \cdot ,N \right\}$, where each user forms a singleton \cite{9632411,8272494}.}

\underline{Step 2: Transfer strategy of each user}~(lines 3--15). Given the current partition ${\Phi}^{(t)} = \{\Phi_1^{(t)},\cdots,\Phi_{J}^{(t)}\}$, every user faces three options: (i) split from the current cluster and work alone by adding LDP noise with scale $\sigma_{\max}$ (i.e., solo training); (ii) split from the current cluster and merge with any other non-empty cluster (if $\alpha_{n,j}\in (0,\alpha_{th})$, a modest amount LDP noise will be added via (\ref{eq:CustomizedLDP})); (iii) stay in the current cluster. The latter two are clustered training strategies. Besides, to prevent the strategic behaviors of participants and social clusters for fairness and partition stability concerns, the following two membership rules for leaving and joining a social cluster are introduced to restrict users' leaving and joining behaviors within each social cluster.
\\
\emph{\textbf{Definition 4} (Membership rule):} The membership rules include the leaving rule and joining rule:
\begin{itemize}
  \item \emph{Leaving rule}: At iteration $t$, if a social cluster $\mathcal{S}\in {\Phi}^{(t)}$ decides to admit a new member, then all its current members cannot leave $\mathcal{S}$ at iteration $t$;
  \item \emph{Joining rule}: At iteration $t$, if any member of a social cluster $\mathcal{S}\in {\Phi}^{(t)}$ leaves, then this social cluster cannot admit any new user at iteration $t$.
\end{itemize}

Next, we define each user's preference order and transferable cluster set.
\\
\emph{\textbf{Definition 5} (Preference order):} The preference order $\succeq_n$ for any user $n\in \mathcal{N}$ is a transitive and complete relation between two transferable social clusters $\mathcal{S}_1$ and $\mathcal{S}_2$ such that:
\begin{align}
\mathcal{S}_1 \succeq_n \mathcal{S}_2 \Leftrightarrow \rho_n(\mathcal{S}_1) \geq \rho_n(\mathcal{S}_2).
\end{align}
Similarly, for the strict preference order $\succ_n$, we have $\mathcal{S}_1 \succ_n \mathcal{S}_2 \Leftrightarrow \rho_n(\mathcal{S}_1) > \rho_n(\mathcal{S}_2)$. Here, $\rho_n(.)$ is the \emph{preference function} for any user $n \in \mathcal{N}$, $n\notin \mathcal{S}$ and any candidate transferable cluster $\mathcal{S} \in {\Phi}^{(t)}$, which is defined as:
\begin{align}\label{preferencefunction}
\rho_n(\mathcal{S}) \!=\!
\begin{cases}
\psi_n(\mathcal{S}\cup\{n\}), & \mathrm{if}\, \psi_l(\mathcal{S}\cup\{n\}) \geq \psi_l(\mathcal{S}),\forall l \in \mathcal{S}, \\
& \&\ \mathcal{S} \notin \mathcal{H}_n~\mathrm{or}~\mathcal{S} \neq \emptyset;\\%\&\ \alpha_{n,j}\!>\!0\
-\infty, & \mathrm{otherwise}.
\end{cases}
\end{align}
$\mathcal{H}_n$ denotes user $n$'s history set which stores the historical clusters that he/her has revisited and been rejected. As any user can always revert to form a singleton, $\mathcal{H}_n$ is only applicable to clusters whose size is greater than one.
\\
\emph{\textbf{Definition 6} (Transferable cluster set):} For each user $n\in \mathcal{S}$, its transferable cluster set at iteration $t$ is defined as:
\begin{align}\label{clusterset}
\mathcal{C}_{n,t}^{\mathrm{cluster}} = \left\{\mathcal{S}' | \rho_n(\mathcal{S}') \geq \rho_n(\mathcal{S}), \forall \mathcal{S}'\in {\Phi}^{(t)}\cup\{\emptyset\} \right\}. %\mathcal{S}' \succeq_n \mathcal{S}
\end{align}

\emph{Remark:}
If user $n$ had been rejected by a cluster $\mathcal{S}'$ (i.e., $\mathcal{S}' \in \mathcal{H}_n$), the cluster $\mathcal{S}'$ will not occur in user $n$'s transferable cluster set.
If $\rho_n(\mathcal{S}') \leq \rho_n(\mathcal{S}), \forall \mathcal{S}'\in {\Phi}^{(t)}\cup\{\emptyset\}$, there is no transferable cluster nor the empty set to transfer for user $n\in \mathcal{S}$, implying that he/she will stay in the current cluster $\mathcal{S}$. Otherwise, user $n$ decides the transfer strategy according to the following transfer rule.
\\
\emph{\textbf{Definition 7} (Transfer rule):} Each user $n\in \mathcal{S}$ sends a merge request to the optimal candidate cluster $\mathcal{S}^{*} \in \mathcal{C}_{n,t}^{\mathrm{cluster}}$ with the largest preference value, i.e.,
$\mathcal{S}^{*} = \operatorname{arg\,max} \rho_n(\mathcal{S}'), \forall \mathcal{S}' \in \mathcal{C}_{n,t}^{\mathrm{cluster}}$.

\emph{Remark:}
If $\mathcal{S}^{*}=\{\emptyset\}$, user $n$ prefers splitting from the current cluster $\mathcal{S}$ and forming a singleton. Otherwise, user $n$ prefers merging with another cluster $\mathcal{S}^{*}$ by splitting from the current cluster $\mathcal{S}$.

\underline{Step 3: Admission strategy of each social cluster} (lines 16--19).
Notably, the transfer order can affect the federal payoffs and the partition result when multiple users ask to join the same cluster $\mathcal{S}$. The following admission rule describes the preference of each cluster for the transfer order.
\\
\emph{\textbf{Definition 8} (Admission rule):} For each social cluster $\mathcal{S}\in {\Phi}^{(t)}$, when it receives multiple transfer requests from multiple candidate users in $\mathcal{C}_{\mathcal{S},t}^{\mathrm{user}}$, it only permits the candidate $n^*$ with the largest preference value, i.e., ${n^*} = \operatorname{arg\,max} \rho_n(\mathcal{S}), \forall n \in \mathcal{C}_{\mathcal{S},t}^{\mathrm{user}}$, and rejects other candidates in $\mathcal{C}_{\mathcal{S},t}^{\mathrm{user}}\backslash \left\{ {{n^*}} \right\}$.

By applying the admission rule, each social cluster $\mathcal{S}\in {\Phi}^{(t)}$ makes its admission strategy. Then, the following split-and-merge operation is executed for each permitted user, which results in a new partition structure, i.e., ${\Phi}^{(t)}\rightarrow{\Phi}^{(t+1)}$.
\\
\emph{\textbf{Definition 9} (Split-and-merge operation):}
A split-and-merge operation that transfers user $n^*\in \mathcal{S}$ to another cluster $\mathcal{S}'$ consists of a split operation (i.e., $\mathcal{S}  \triangleright\big\{ {\mathcal{S}^-,\left\{ {{n^*}} \right\}} \big\}$) and a subsequent merge operation (i.e., $\big\{ {\mathcal{S}',\left\{ {{n^*}} \right\}} \big\}\triangleright \mathcal{S}'^+$). Here, $\mathcal{S}^- = \mathcal{S}\backslash \left\{ {{n^*}} \right\}$ and $\mathcal{S}'^+ = \mathcal{S}' \cup \left\{ {{n^*}} \right\}$.

The above steps 2-3 end until reaching a final Nash-stable partition structure $\Phi^*$ (line 2).
\\
\emph{\textbf{Definition 10} (Nash-stability):} A partition $\Phi$ is Nash-stable if $\mathcal{S} \succeq_n \mathcal{S}', \forall n \in \mathcal{S} \subseteq \mathcal{N}, \forall \mathcal{S}' \in \Phi\cup \{\emptyset\}$.
\\
\emph{\textbf{Theorem 2}:}
The partition outcome $\Phi^*$ in Alg.~\ref{Algorithm1} is Nash-stable.
%\end{theorem}
\begin{IEEEproof}
We first prove that Alg.~\ref{Algorithm1} can always converge to a final disjoint partition ${\Phi}^*$, given an arbitrary initial structure $\Phi^{(0)}$. By inspecting the preference function in (\ref{preferencefunction}), we can observe that each single split-and-merge operation either results in: (i) an unvisited new partition; or (ii) a singleton with a non-cooperative user.
For case (i), as the maximum number of partitions among users in $\mathcal{N}$ is finite and can be obtained by the well-known Bell number function \cite{7817887}, the number of transformations in $\{{\Phi ^{(0)}} \to  \cdots  \to {\Phi ^{(t)}} \to  \cdots  \to {\Phi ^{(T)}}={\Phi}^{*}\}$ is finite. For case (ii), in the next iteration $t+1$, the non-cooperative user should either remain non-cooperative or join a new cluster (which yields an unvisited partition). In all cases, the transformation sequence will terminate after $T$ turns and converge to a final outcome ${\Phi}^{*}$.

Next, we prove the Nash-stability by contradiction. Assume that the final partition $\Phi^*$ in Alg.~\ref{Algorithm1} is not Nash-stable. As such, there is a user $n \in \mathcal{S}$ and a cluster $\mathcal{S}' \in \Phi^*\cup \{\emptyset \}, \mathcal{S}'\ne \mathcal{S}$ such that $\mathcal{S}' \succeq_n \mathcal{S}$. Thereby, user $n$ will split from the current cluster $\mathcal{S}$ and merge with $\mathcal{S}'$, contradicting with the fact that $\Phi^*$ is the convergence outcome of Alg.~\ref{Algorithm1}.
\end{IEEEproof}

\emph{Remark:}
Theorem 2 indicates that any final partition derived from Alg.~\ref{Algorithm1} is Nash-stable and individually rational. Namely, no user $n\in \mathcal{N}$ tends to leave the current cluster $\Phi_j^*$ and switch to another cluster $\Phi_l^* \in \Phi^*\cup \{\emptyset\}$, $j\ne l$ to improve his/her individual payoff. {Notably, the Nash-stable partition outcome ${\Phi}^{*}$ produced by Alg.~\ref{Algorithm1} is not unique. For example, different initial partitions may result in distinct partition outcomes.}
The overall computational complexity of Alg.~\ref{Algorithm1} is $\mathcal{O}( T\cdot N\cdot {\left| {\Phi^{i}} \right|})$ in the worst case.
Besides, simulation results from Fig.~\ref{fig:simu11} and Fig.~\ref{fig:simu13} show that our proposed Alg.~\ref{Algorithm1} can quickly converge to the Nash-stable partitions. It indicates that our SCFL only incurs small additional overheads in the social cluster changes (i.e., additions or subtractions) by users to ensure the availability of FL models.

\section{EXPERIMENTAL VALIDATION}\label{sec:SIMULATION}
In this section, we conduct extensive experiments using the real Facebook social network and classic MNIST/CIFAR-10 dataset on a workstation with Intel Xeon Platinum 8280 CPU (2.7GHz/4.0GHz), 256G RAM, and two Nvidia GeForce RTX 3090 GPUs. We use PyTorch to implement the SCFL.

\vspace{-3mm}
\subsection{Experiment Setup}\label{subsec:evalution1}
\begin{figure*}[htbp]%\vspace{-0.15cm}
\begin{minipage}[t]{0.315\textwidth}
\centering
    \includegraphics[height=4cm,width=0.98\textwidth]{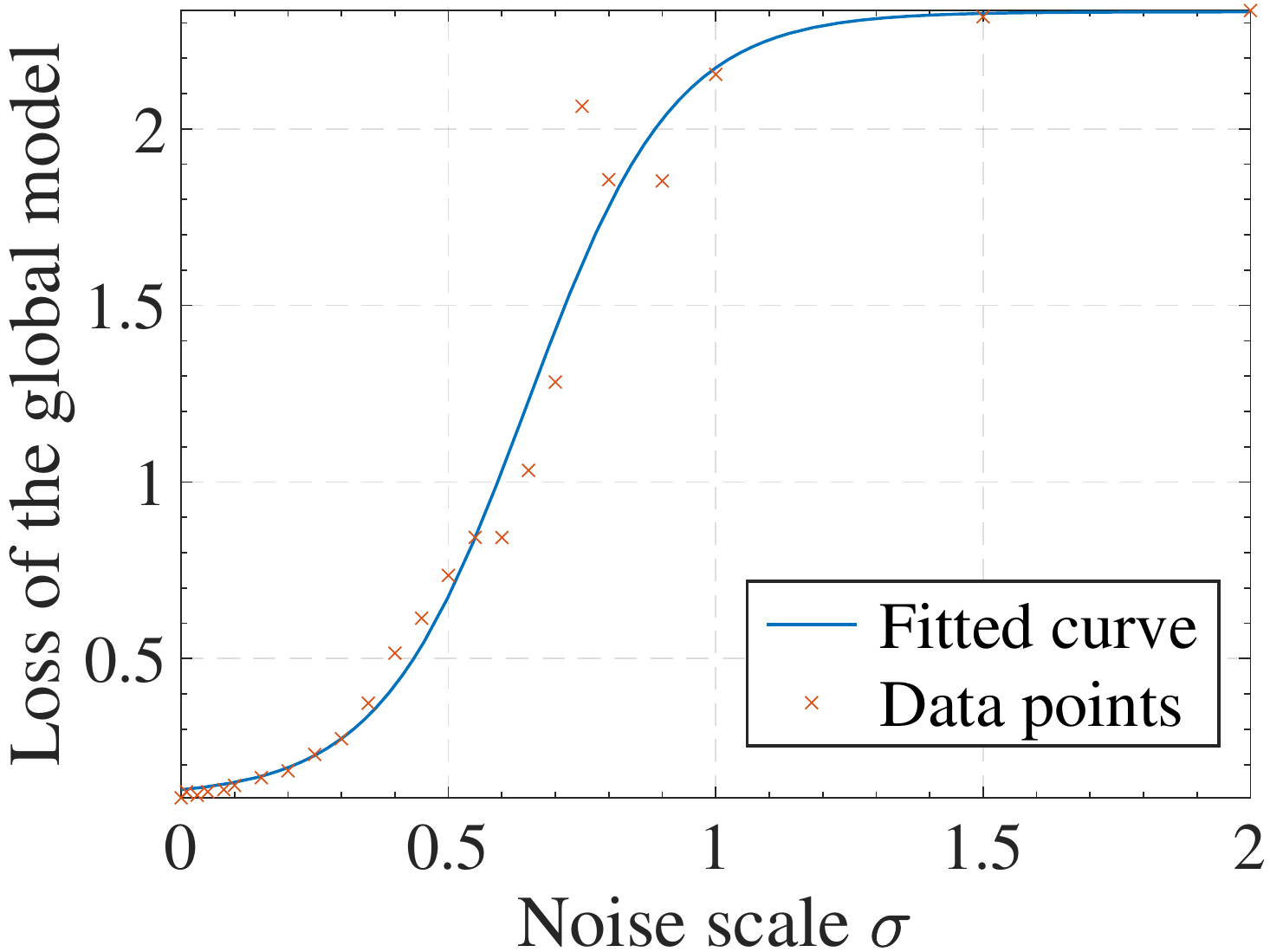}
    \caption{Curve fitting of loss ${\mathscr{L}}$ w.r.t noise scale $\sigma$ under IID: ${\mathscr{L}} = \frac{0.0225}{0.010+\exp{(-6.9672\sigma)}}+0.09$.}\label{fig:simu1}
\end{minipage}~~
\begin{minipage}[t]{0.32\textwidth}
\centering
    \includegraphics[height=4cm,width=0.98\textwidth]{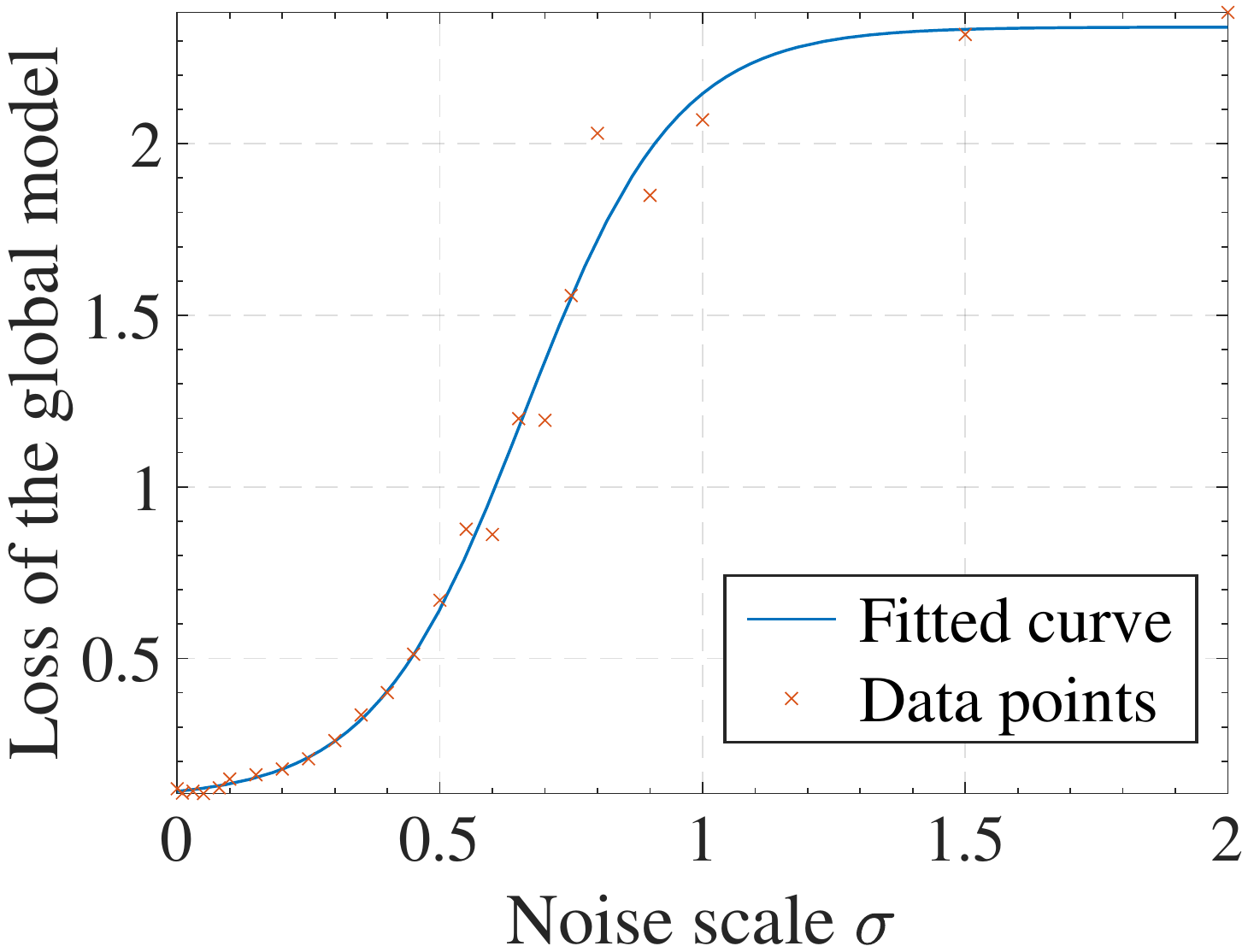}
    \caption{Curve fitting of loss w.r.t noise scale in non-IID ($\gamma\!=\!0.6$): ${\mathscr{L}} \!=\! \frac{0.02}{0.009+\exp{(-7.278\sigma)}}\!+\!0.109$.}\label{fig:simu2}
\end{minipage}~~
\begin{minipage}[t]{0.32\textwidth}
\centering
    \includegraphics[height=4cm,width=0.98\textwidth]{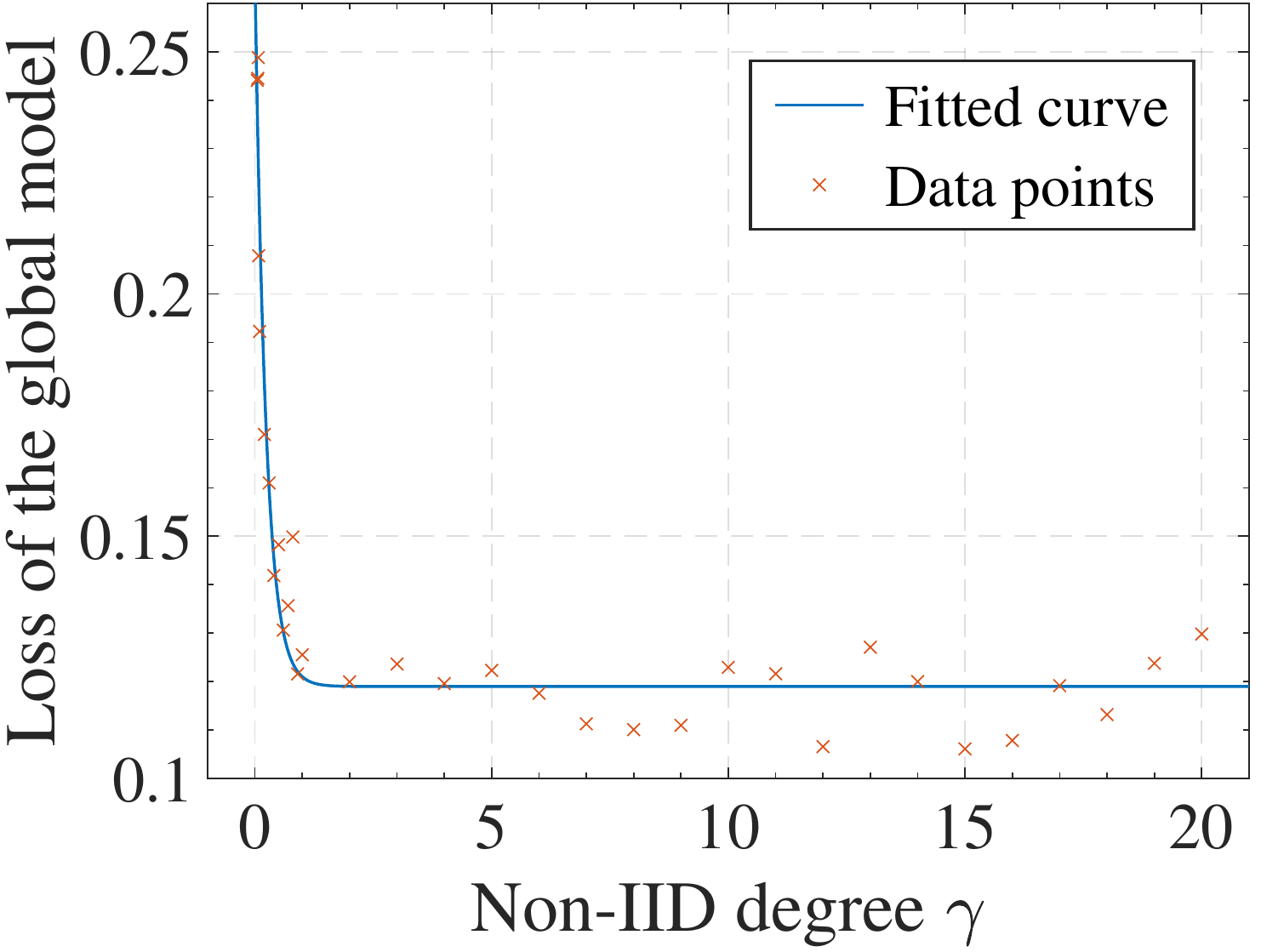}
    \caption{Curve fitting of loss w.r.t non-IID degree $\gamma$ with $\sigma\!=\!0.1$: ${\mathscr{L}}\! =\! 0.147\exp{(-4.288\gamma)}\!+\!0.119$.}\label{fig:simu3}
\end{minipage}%\vspace{-3.5mm}
\end{figure*}

\begin{figure*}[htbp]%\vspace{-0.15cm}
\begin{minipage}[t]{0.315\textwidth}
\centering
    \includegraphics[height=4cm,width=0.99\textwidth]{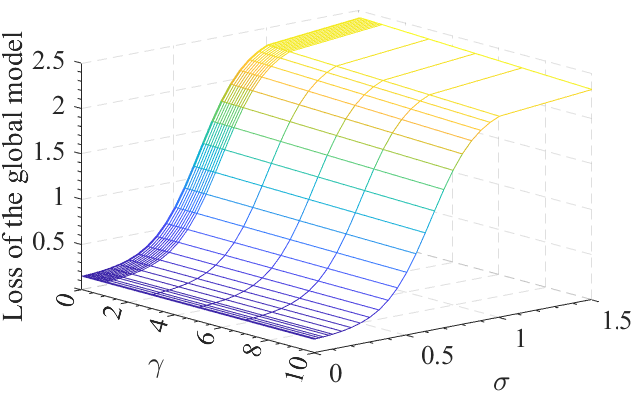}
    \caption{Curve fitting of the loss function ${\mathscr{L}}(\sigma,\gamma)$ in (\ref{eq:lossfitting}) w.r.t noise scale $\sigma$ and non-IID degree $\gamma$ {in MNIST using the NLLLoss loss function}: ${\mathscr{L}}\left( \sigma ,\gamma \right) = \frac{0.013\exp{(-0.0044\gamma)}}{0.0057+\exp{(-8.18\sigma)}}+0.14$.}\label{fig:simu4}
\end{minipage}~~
\begin{minipage}[t]{0.32\textwidth}
\centering
    \includegraphics[height=4cm,width=0.99\textwidth]{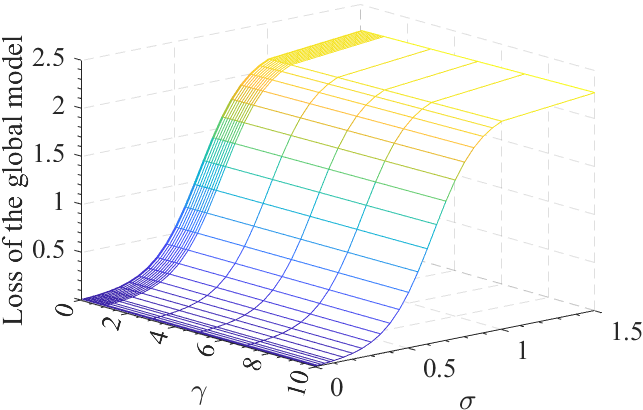}%\vspace{-0.4cm}
    {\caption{Curve fitting of the loss function ${\mathscr{L}}(\sigma,\gamma)$ in (\ref{eq:lossfitting}) w.r.t noise scale $\sigma$ and non-IID degree $\gamma$ in MNIST using the MSE loss function: $\mathscr{L}\left( \sigma ,\gamma \right) =\frac{0.013\exp{(-0.0021\gamma)}}{0.0057+\exp{(-8.20\sigma)}}+0.14$.}\label{fig:simu4-2}}
\end{minipage}~~
\begin{minipage}[t]{0.32\textwidth}
\centering
    \includegraphics[height=4cm,width=1\textwidth]{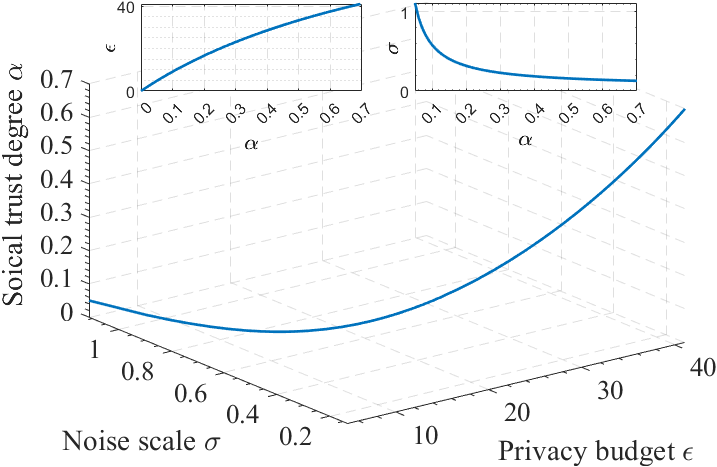}%\vspace{-0.4cm}
    {\caption{The relationship between social trust degree $\alpha$, privacy budget $\epsilon$, and noise scale $\sigma$ in MNIST when $\theta_1 =100$ and $\theta_2 =1$.}\label{fig:simu4-3}}
\end{minipage}\vspace{-2.5mm}
\end{figure*}

\textbf{Datasets and Models.} We evaluate SCFL on the Facebook ego network \cite{FaceBookEgo} with 4039 nodes (i.e., social users) and over 88K edges (i.e., social relations), which shows a real social network topology. The participants are randomly chosen from the Facebook network with varying numbers, i.e., $[50,100,150,200,250]$, where their social connections are extracted from the Facebook network {(i.e., whether there exists an edge between them).}
{As the Facebook ego network does not offers the social strength of these edges, we synthesize the social relationship $e_{n,m}$ of socially connected users via a truncated normal distribution as in \cite{9119841}. The asymmetric social relationship is possible, i.e., $e_{n,m} \ne e_{m,n}$.}
Two typical datasets for FL tasks are considered, i.e., the MNIST dataset\footnotemark[10] for handwritten digits recognition and the CIFAR-10 dataset\footnotemark[11] for image recognition. \footnotetext[10]{http://yann.lecun.com/exdb/mnist}\footnotetext[11]{https://www.cs.toronto.edu/~kriz/cifar.html}
For dataset partition among individuals, the non-IID degree of users' local dataset is controlled by varying the Dirichlet parameter $\gamma$ (as analyzed in Sect.~\ref{subsec:scheme4}). The value of $\gamma$ is selected between $[0.05,20]$.
For local model training, the 4-layer CNN model is applied for MNIST, while the 5-layer CNN model is adopted for CIFAR-10. The total numbers of communication rounds are set as $30$ and $100$ in MNIST and CIFAR-10, respectively. For both MNIST and CIFAR-10, the mini-batch SGD with learning rate $\eta=0.05$, local batch size $64$, and local epoch $1$ is adopted for all users. {For the same FL task, the hyperparameters including communication rounds, learning rate, batch size, and number of participants are same.}% \cite{726791}\cite{CIFAR10}

\textbf{LDP Noise Adding.} For customized LDP perturbation, we set $\alpha_{th}=0.7$, $\theta_1 =100$ (MNIST), $\theta_1 =100$ (CIFAR10), $\theta_2 = 1$, and $\delta=10^{-6}$ in (\ref{eq:CustomizedLDP}) to map the social trust $\alpha_{n,j}$ to the privacy protection level $\epsilon_{n,j}$ for privacy-utility tradeoff under the clustered training strategy. We set $\sigma_{n,j} = \frac{\sqrt{2\log(1.25/\delta)}}{\epsilon_{n,j}}$ in (\ref{eq:noisescale}) based on \cite{9820771,Abadi2016Deep}. The Gaussian noise scale under the solo training strategy is set as $\sigma_{\max}=0.6$ for MNIST and $\sigma_{\max}=0.3$ for CIFAR-10, respectively.

\textbf{Federation Game.}
For federation game model, we set $\omega=0.8$, $\varsigma=30$, $\lambda_p=0.52$, $\lambda_c=1.2$, $\kappa_1=35.4278$, $\kappa_2=102.2444$.
To evaluate the effect of social attributes in SCFL, for socially connected users in the real Facebook ego network, we further set the following three levels of social effects. For \emph{strong social effects}, the mutual social trust values between socially connected users are larger than the threshold $\alpha_{th}$; while for \emph{weak social effects}, users' social trust values are randomly distributed within $[0,\alpha_{th}]$. For \emph{no social effects}, there exist no social connections among users, and our strategy automatically degenerates to the typical cross-device FL setting with non-cooperative users.

\begin{figure*}[htbp]%\vspace{-0.2cm}
\setlength{\abovecaptionskip}{-0.0cm}
\begin{minipage}[t]{0.32\textwidth}
\centering
    \includegraphics[height=4cm,width=0.98\textwidth]{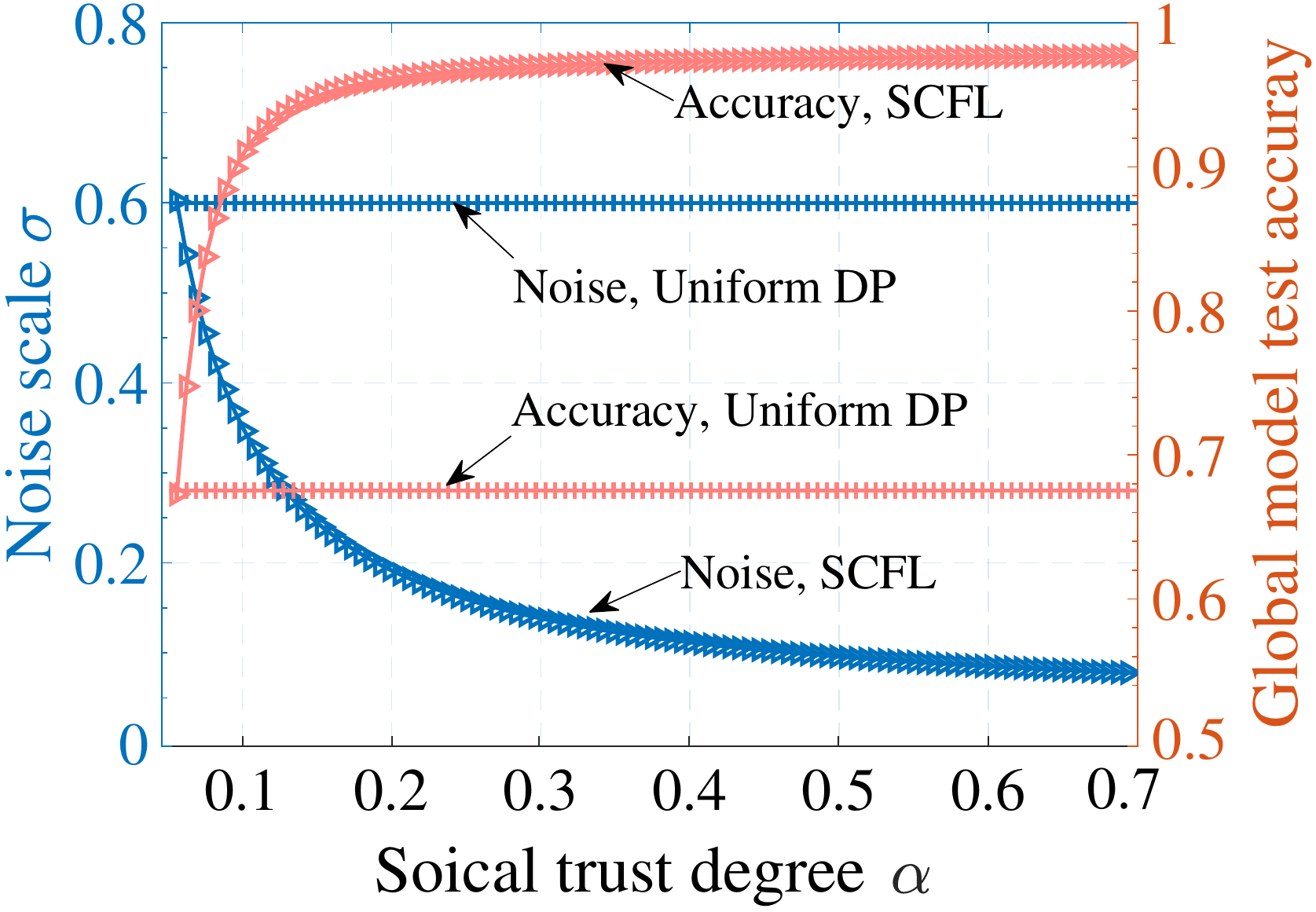}%width=\textwidth
    {\caption{Effect of customized local perturbation in terms of noise scale $\sigma$ and global model test accuracy vs. social trust degree in MNIST.}\label{fig:simu5}}
\end{minipage}~~
\begin{minipage}[t]{0.32\textwidth}
\centering
    \includegraphics[height=4cm,width=0.95\textwidth]{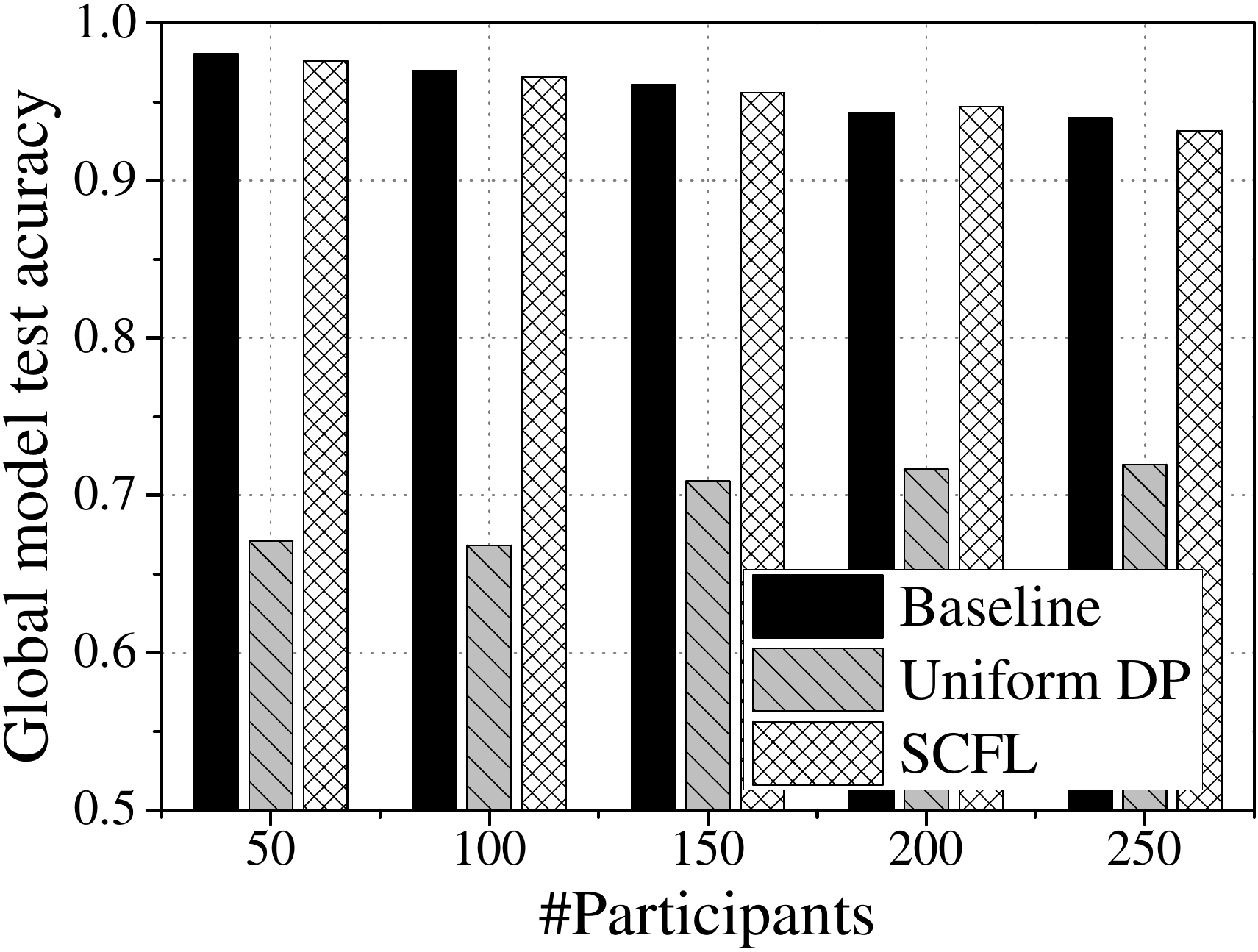}
    \caption{Comparison of global model test accuracy in MNIST in three schemes under different numbers of participants.}\label{fig:simu6}
\end{minipage}~~
\begin{minipage}[t]{0.32\textwidth}
\centering
    \includegraphics[height=4cm,width=0.98\textwidth]{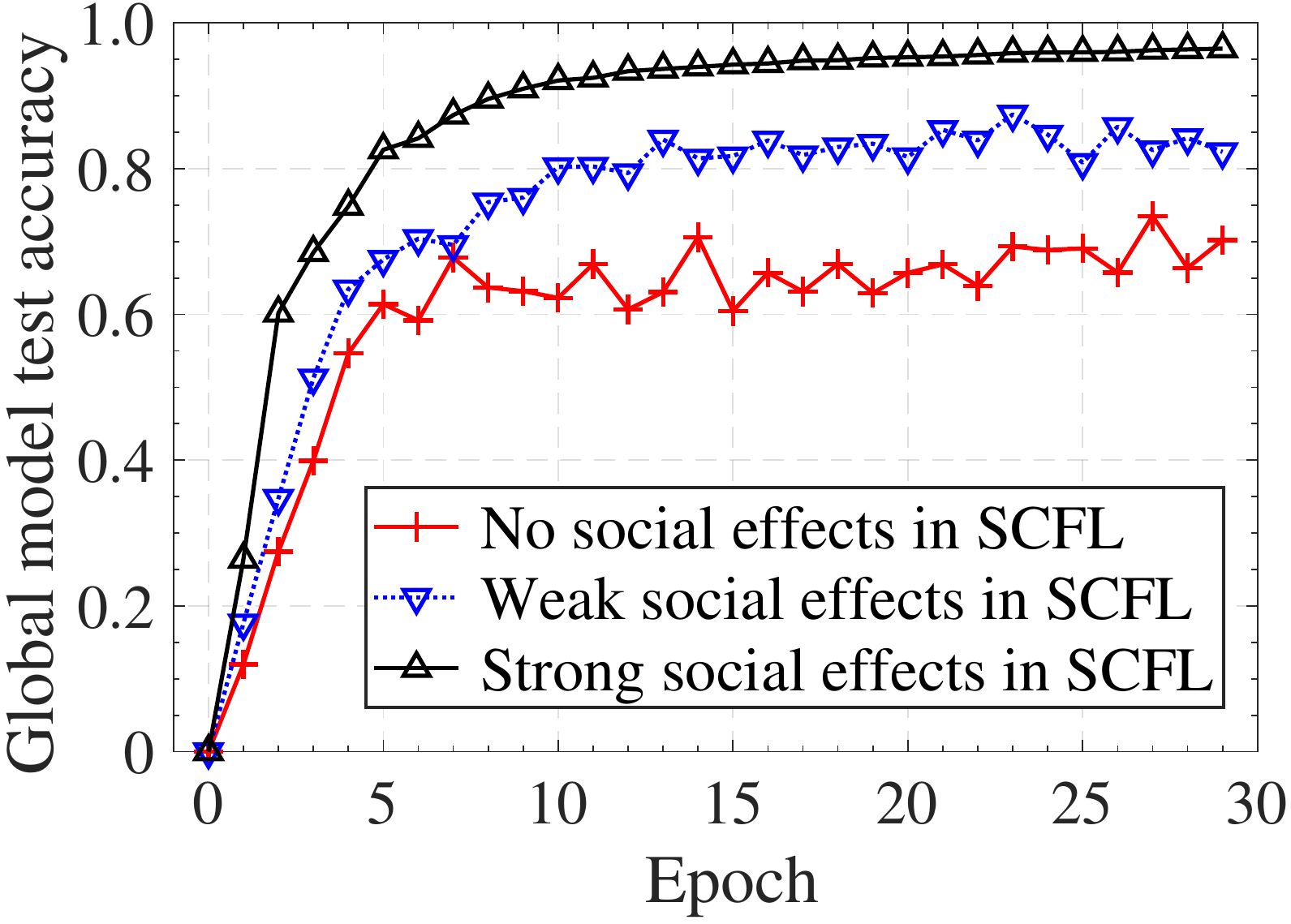}
    \caption{Evolution of global model test accuracy in MNIST under different levels of social effects in SCFL ($N=100$).}\label{fig:simu7}
\end{minipage}%\vspace{-0.1cm}
\end{figure*}

\begin{figure*}[htbp]%\vspace{-0.2cm}
\setlength{\abovecaptionskip}{-0.0cm}
\begin{minipage}[t]{0.32\textwidth}
\centering
    \includegraphics[height=4cm,width=0.98\textwidth]{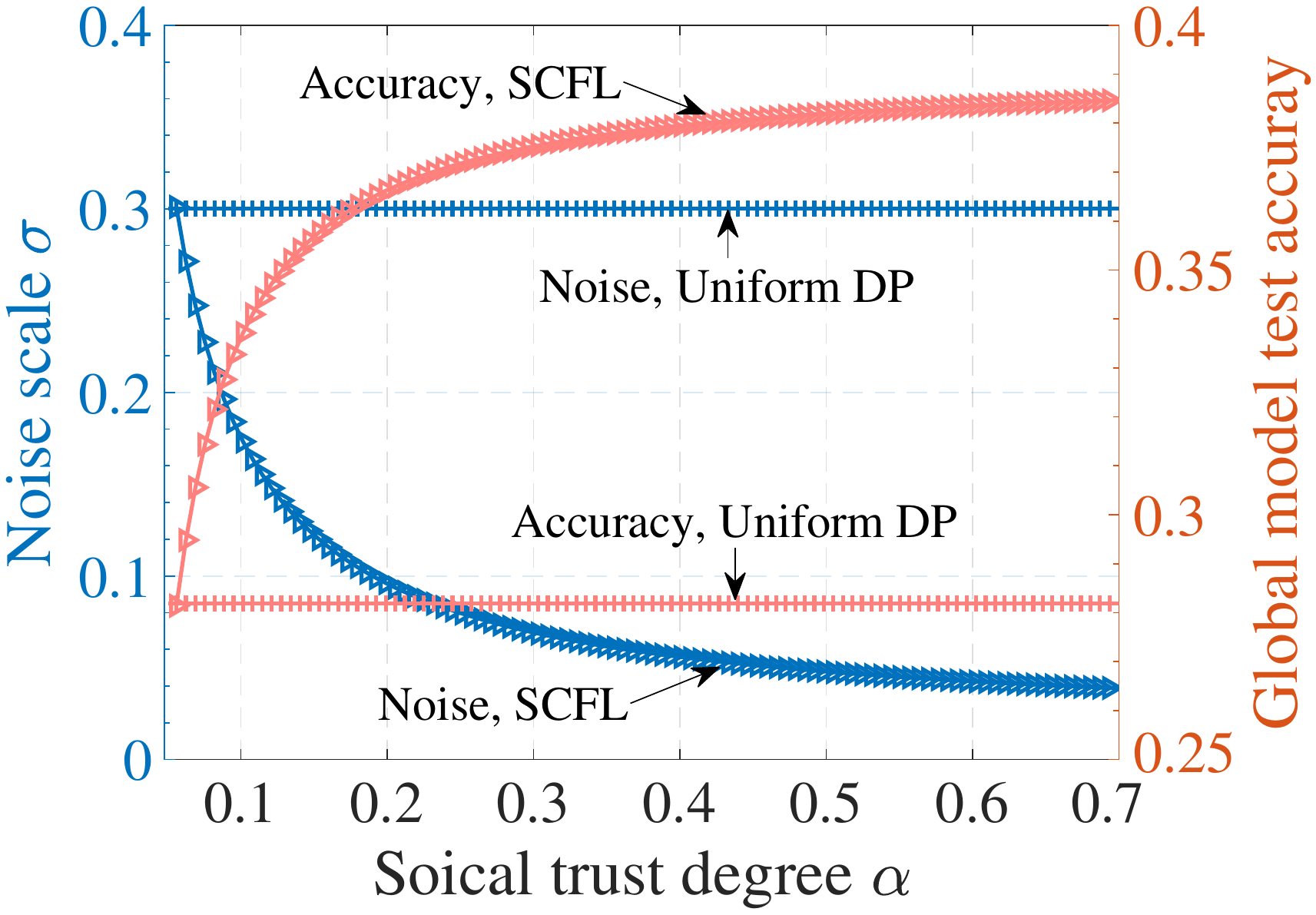}
    \caption{Effect of customized local perturbation in terms of noise scale $\sigma$ and global model test accuracy vs. social trust degree in CIFAR-10.}\label{fig:simu8}
\end{minipage}~~
\begin{minipage}[t]{0.32\textwidth}
\centering
    \includegraphics[height=4cm,width=0.95\textwidth]{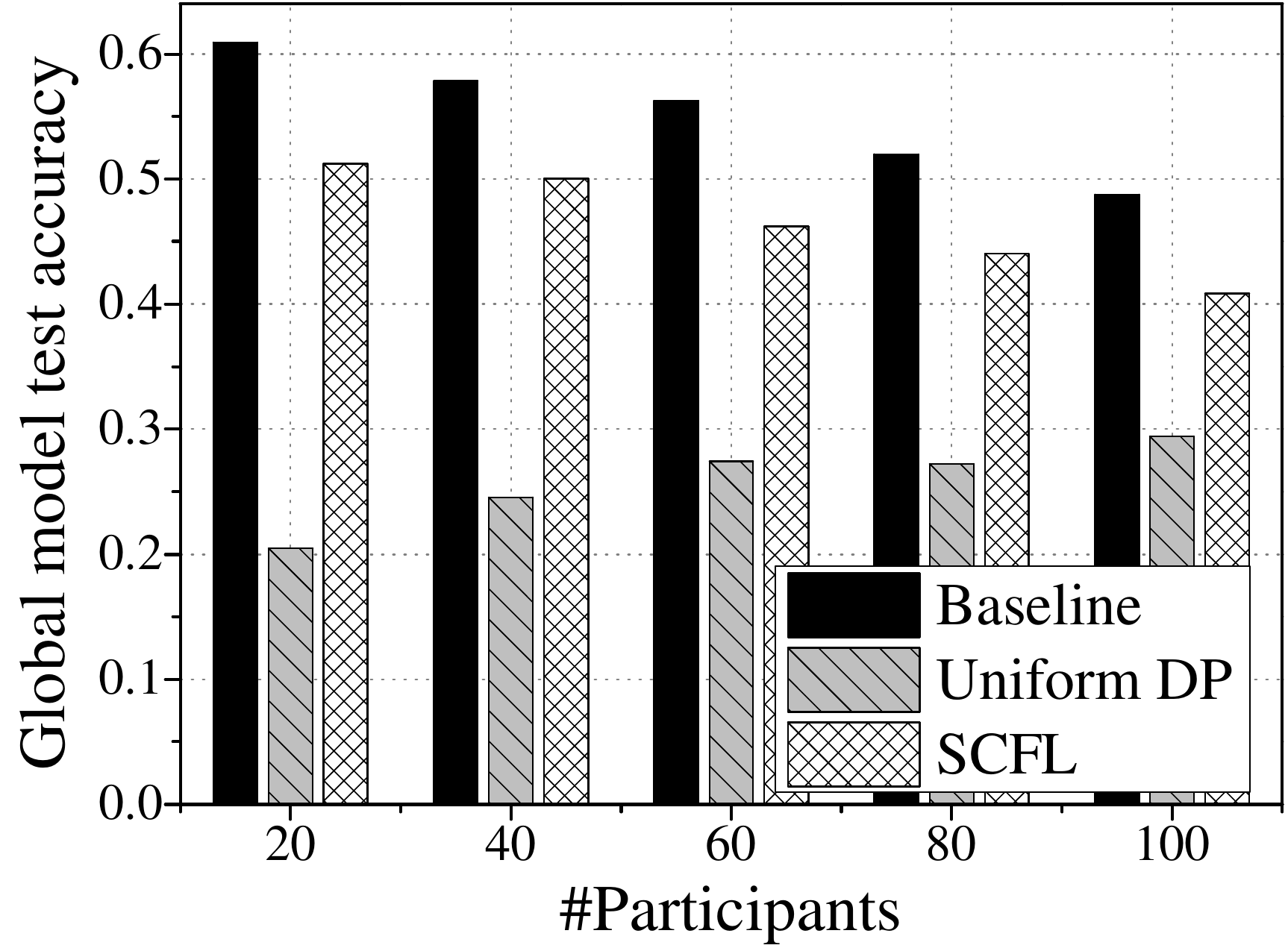}
    \caption{Comparison of global model test accuracy in CIFAR-10 in three schemes under different numbers of participants.}\label{fig:simu9}
\end{minipage}~~
\begin{minipage}[t]{0.32\textwidth}
\centering
    \includegraphics[height=4cm,width=0.98\textwidth]{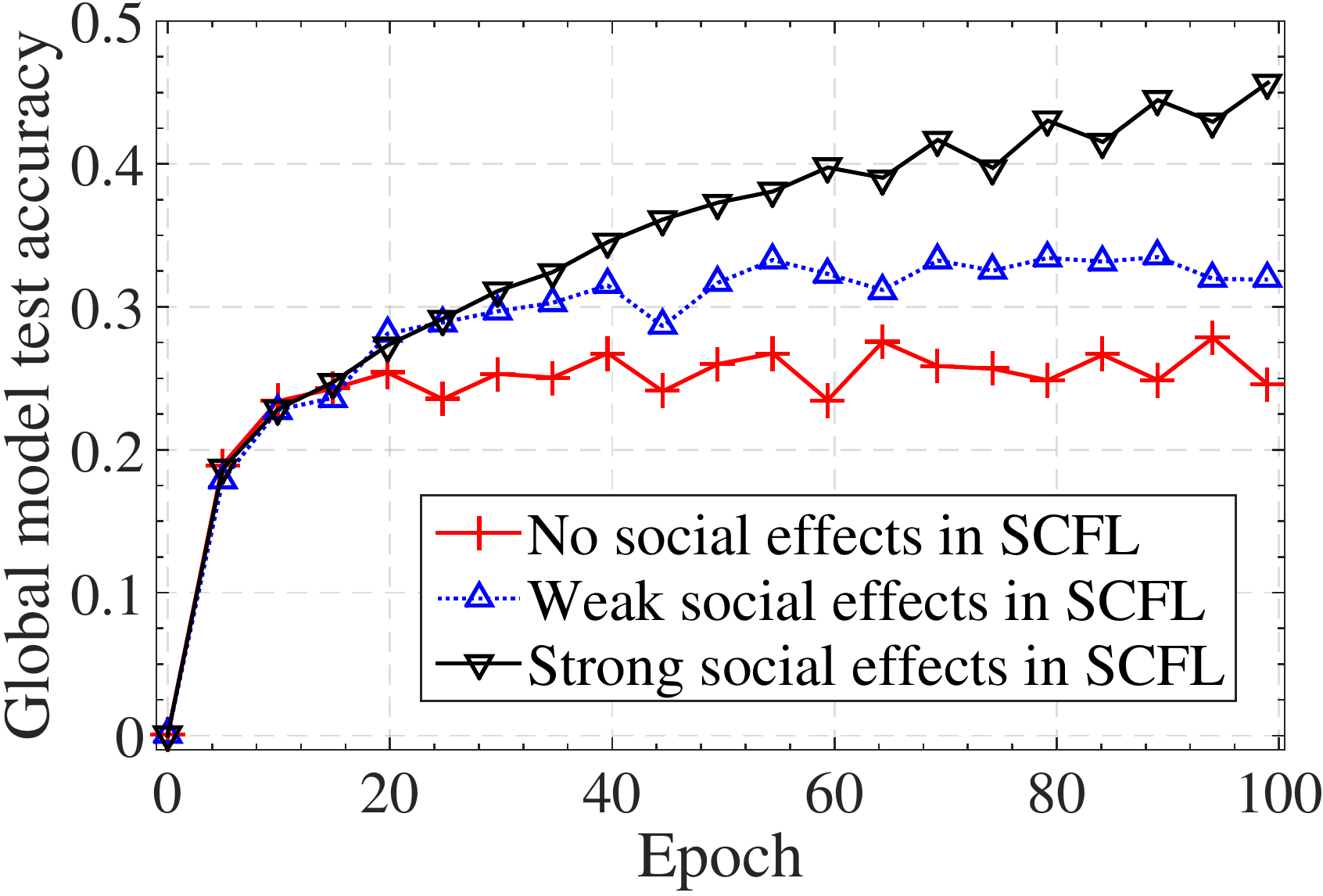}
    \caption{Evolution of global model test accuracy in CIFAR-10 under different levels of social effects in SCFL ($N=100$).}\label{fig:simu10}
\end{minipage}%\vspace{-0.1cm}
\end{figure*}

%\textbf{Comparing Methods.}
The performance of SCFL is evaluated by comparing with the following conventional schemes.
\begin{itemize}
  \item \textbf{Uniform DP scheme \cite{9410383}}. In \cite{9410383}, the local model updates of social users are sanitized by adding LDP noise with the uniform scale before global aggregation. As users usually have distinct privacy expectations, the relatively large LDP noise (i.e., $\sigma_{\max}$) is applied in practice to meet the requirements of most users. Besides, we set $\gamma=0.6$ for MNIST and $\gamma=1$ for CIFAR-10.
  \item \textbf{Non-cooperative scheme}. In most works on LDP-based FL such as \cite{9820771,9740410}, individuals apply the solo training strategy and act as singletons in conducting FL tasks.
  \item \textbf{Social influence based scheme}. In this scheme, the top-$K$ nodes with the highest social influence invite others to form disjoint social clusters, and users can dynamically transfer across these clusters. Here, we set $K=10$.
\end{itemize}

\vspace{-4mm}
\subsection{Verification for Model Utility Function}\label{subsec:evalution2}
In Figs.~\ref{fig:simu1}--\ref{fig:simu4}, we verify the model utility function in FL (measured by the loss of trained model) defined in (\ref{eq:lossfitting}) on MNIST dataset by training the 4-layer CNN model.
Here, the number of participants (i.e., $N$) is set as $100$. As shown in Figs.~\ref{fig:simu1} and \ref{fig:simu2}, the relationship between model loss and Gaussian noise scale under both IID and non-IID cases can be fitted as a sigmoid curve; while Fig.~\ref{fig:simu3} shows that the relationship between model loss and non-IID degree can be fitted as an exponential function.
{In Fig.~\ref{fig:simu4} and Fig.~\ref{fig:simu4-2}, we evaluate the loss of the trained model on two loss functions: typical negative log likelihood loss (NLLLoss) and the mean-square error (MSE), respectively.}
In Fig.~\ref{fig:simu4}, by varying both noise scale and non-IID degree, the model loss function can be well-fitted by the 3D sigmoid curve with curve fitting parameters $\mu_1=0.013$, $\mu_2=0.0044$, $\mu_3=0.0057$, $\mu_4=8.18$, and $\mu_5=0.14$.
{As seen in Fig.~\ref{fig:simu4-2}, the loss can still be well-fitted by 3D sigmoid curve, and the curve fitting parameters are $\mu_1=0.013$, $\mu_2=0.0021$, $\mu_3=0.0057$, $\mu_4=8.20$, and $\mu_5=0.14$.
From Fig.~\ref{fig:simu4} and Fig.~\ref{fig:simu4-2}, we can observe} that the effect of LDP noise overwhelms the non-IID effect in model utility degradation.
{Besides, in Fig.~\ref{fig:simu4-3}, it can be observed that given $\theta_1 =100$ and $\theta_2 =1$, when $\alpha \in [0,\alpha _{th}]$, we have $\epsilon \in [ 0,40]$; and when $\alpha \in [0.05,\alpha _{th}]$, we have $\sigma \in [0, 1.1]$. Under this setting, the customizable privacy budget $\epsilon$ is within a certain range, so that the added DP noise $\sigma$ will not destroy the model performance.}

\vspace{-3mm}
\subsection{Numerical Results}
\begin{figure*}[htbp]%\vspace{-0.2cm}
\setlength{\abovecaptionskip}{-0.0cm}
\begin{minipage}[t]{0.32\textwidth}
\centering
    \includegraphics[height=4.1cm,width=0.98\textwidth]{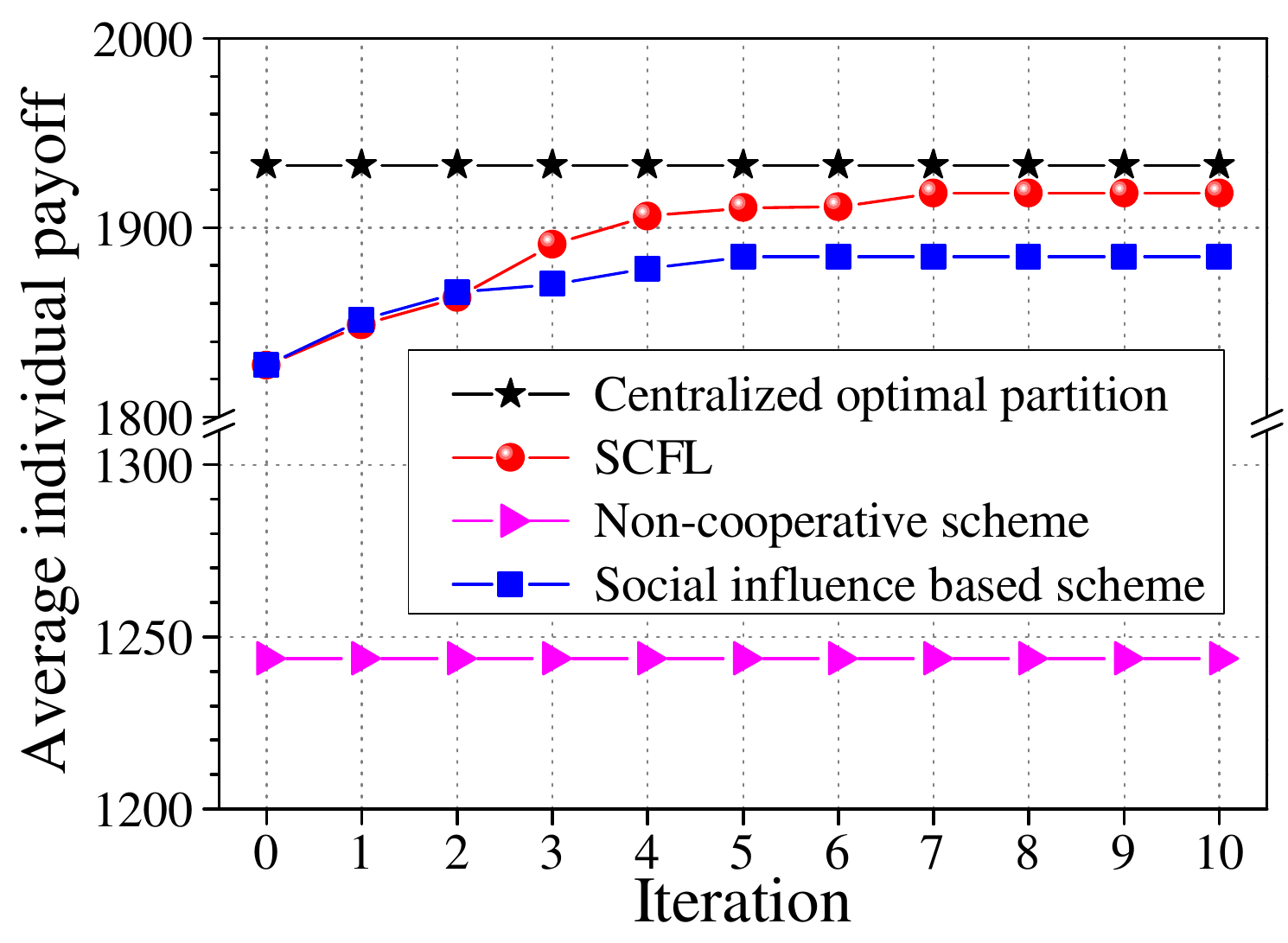}
    \caption{Evolution of average individual payoff in SCFL, compared with other three conventional schemes ($N=100$).}\label{fig:simu11}
\end{minipage}~~
\begin{minipage}[t]{0.32\textwidth}
\centering
    \includegraphics[height=4.1cm,width=0.95\textwidth]{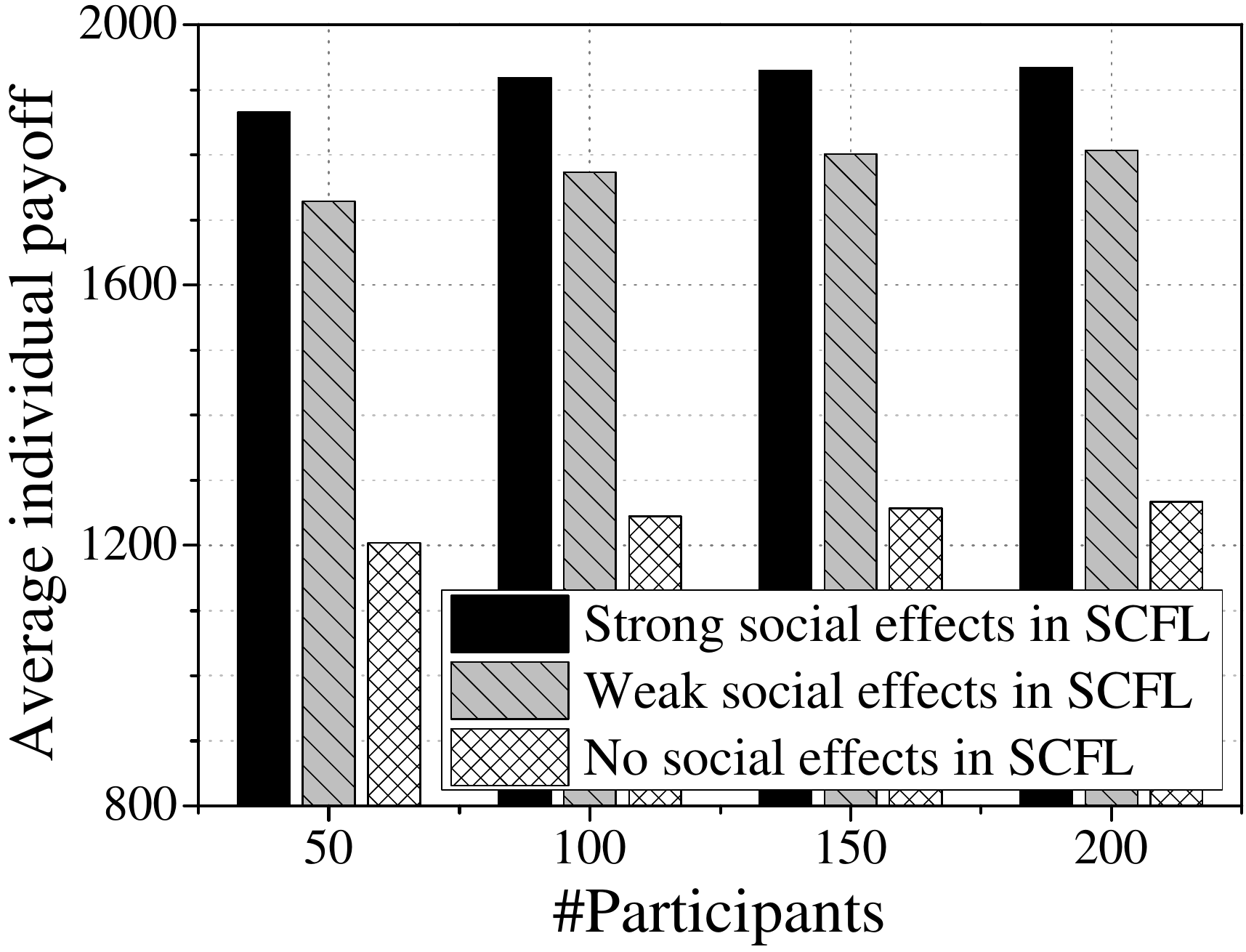}
    \caption{Comparison of average individual payoff in SCFL under different numbers of participants and different levels of social effects.}\label{fig:simu12}
\end{minipage}~~
\begin{minipage}[t]{0.32\textwidth}
\centering
    \includegraphics[height=4.1cm,width=0.95\textwidth]{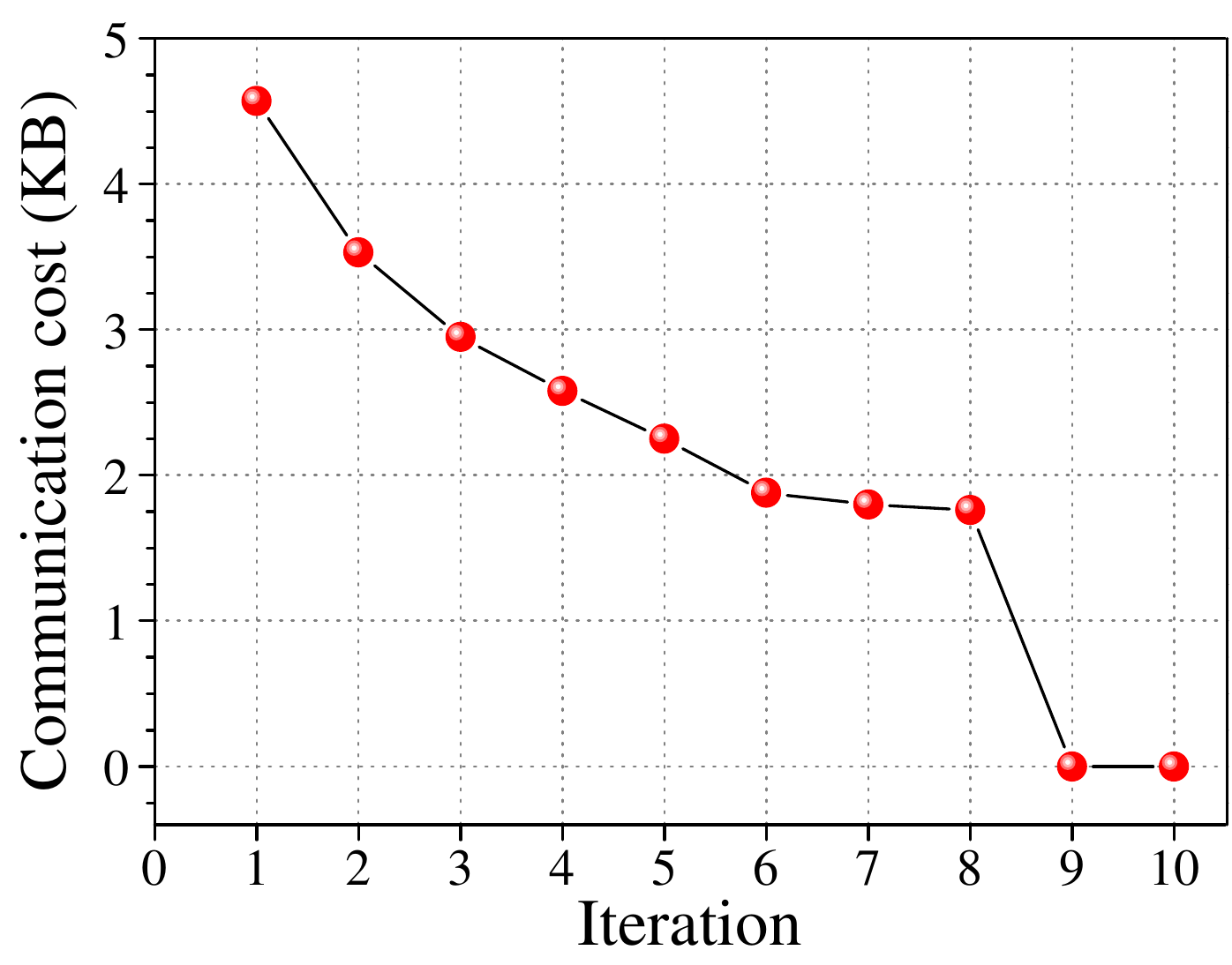}
    {\caption{Evolution of communication cost in forming a Nash-stable partition of social clusters in SCFL ($N=100$).}\label{fig:simu14}}
\end{minipage}%\vspace{-0.2cm}
\end{figure*}

\begin{figure*}[htbp]
\setlength{\abovecaptionskip}{-0.0cm}
\begin{minipage}[t]{0.32\textwidth}
\centering
    \includegraphics[height=4.1cm,width=1\textwidth]{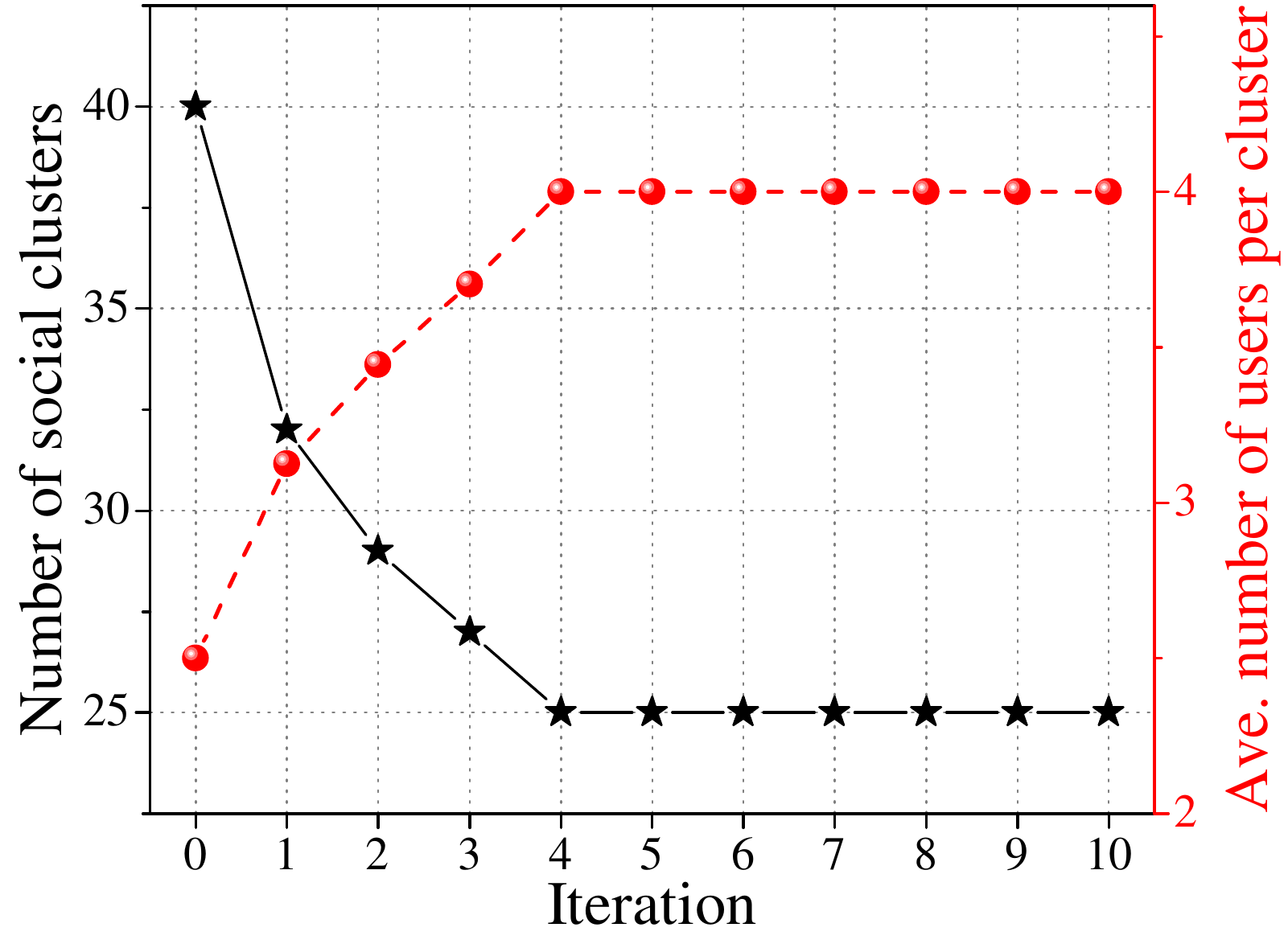}
    \caption{Evaluation of social clustering results of the FTU game in terms of number of social clusters and average number of users per cluster ($N=100$).}\label{fig:simu13}
\end{minipage}~~
\begin{minipage}[t]{0.32\textwidth}
\centering
    \includegraphics[height=4.1cm,width=1.0\textwidth]{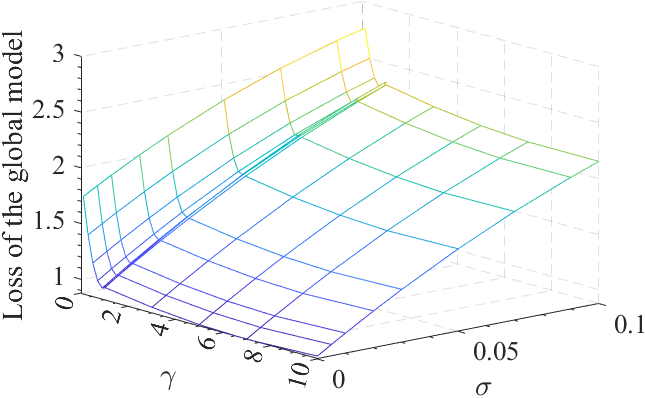}%\vspace{-0.38cm}
    {\caption{Curve fitting of the loss ${\mathscr{L}}$ w.r.t noise scale $\sigma$ and non-IID degree $\gamma$ in the CIFAR10 dataset using the Resnet18 model.}\label{fig:simu4-4}}
\end{minipage}~~
\begin{minipage}[t]{0.32\textwidth}
\centering
    \includegraphics[height=4.1cm,width=1\textwidth]{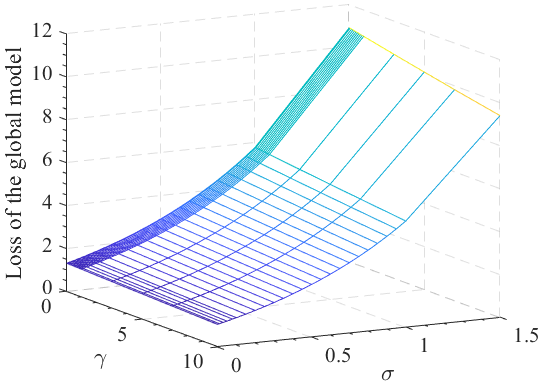}%\vspace{-0.38cm}
    {\caption{Curve fitting of loss ${\mathscr{L}}$ w.r.t noise scale $\sigma$ and non-IID degree $\gamma$ in the 20newsgroup dataset using the MLP model.}\label{fig:simu4-5}}
\end{minipage}\vspace{-0.2cm}
\end{figure*}

Using Figs.~\ref{fig:simu5}--\ref{fig:simu10}, we first evaluate the effects of customized local perturbation, number of participants, and level of social effects in SCFL on MNIST and CIFAR-10 datasets.
Next, we analyze the user payoff and stability of the federation game in SCFL using Figs.~\ref{fig:simu11}--\ref{fig:simu13}, by comparing with existing schemes.
After that, we validate the feasibility of SCFL in complex models and language tasks using Figs.~\ref{fig:simu4-4}--\ref{fig:simu4-5}.

\textbf{Effect of Customized Local Perturbation.} Figs.~\ref{fig:simu5} and \ref{fig:simu8} illustrate the effects of the average social trust degrees in terms of Gaussian noise scale and global model test accuracy on the MNIST dataset and CIFAR-10 dataset, respectively.
As seen in Figs.~\ref{fig:simu5} and \ref{fig:simu8}, a higher social trust degree $\alpha$ in SCFL results in a lower scale of added Gaussian noise and correspondingly higher model accuracy on both MNIST and CIFAR-10, thereby enforcing customizable privacy protection. While in the uniform DP scheme, as the uniform and relatively large Gaussian noise is applied for all users, the global model accuracy keeps unvaried and at a low level under different $\alpha$.

\textbf{Effect of Number of Participants.} Figs.~\ref{fig:simu6} and \ref{fig:simu9} compare the FL model performance in three schemes on MNIST and CIFAR-10, respectively. In the baseline scheme (i.e., naive FL under IID without DP perturbation), participants' raw local updates rather than the noised version are exchanged, where the social clustering, non-IID efects, and user privacy preservation are not considered.
As shown in Figs.~\ref{fig:simu6} and \ref{fig:simu9}, the SCFL outperforms the uniform DP scheme in attaining a smaller accuracy gap with the baseline scheme in both MNIST and CIFAR-10 datasets. Besides, when the number of participants increases, the model accuracy in both the SCFL and the baseline scheme decreases, while that in the uniform DP scheme increases. The reason is that in our setting, the total dataset is divided among all participants in an non-IID manner.
As such, more participants result in lower local samples of each user, causing an accuracy drop in our SCFL and the baseline scheme. Moreover, when more participants add the random noise with the same Gaussian distribution, the aggregated effect of LDP noise can be reduced, thereby causing an accuracy rise in the uniform DP scheme.

\textbf{Level of Social Effects.} Figs.~\ref{fig:simu7} and \ref{fig:simu10} depict the evolution of global model test accuracy in our SCFL under different levels of social effects in MNIST and CIFAR-10 datasets, respectively.
As observed in Figs.~\ref{fig:simu7} and \ref{fig:simu10}, the test accuracy of global model in FL under strong social effects is higher than that under weak or no social effects on both MNIST and CIFAR-10, validating the effects of social attributes in combination with FL.

%\vspace{-0.5mm}
\textbf{Individual Payoff.} Fig.~\ref{fig:simu11} compares the average individual payoff in our SCFL, compared with other three conventional schemes.
In Fig.~\ref{fig:simu11}, it can be seen that the SCFL fast converges after $7$ iterations and yields a near-optimal performance compared with the centralized optimal partition, which is superior to the non-cooperative scheme and social influence based scheme. It is because the partition structure with fixed cluster heads in the social influence based scheme cannot adapt to heterogeneous users and the varying FL environment, leading to a smaller individual payoff. In the non-cooperative scheme, as all users work alone in the FL process, it causes the lowest individual payoff.
Fig.~\ref{fig:simu12} shows the average individual payoff in our SCFL under different numbers of participants and various levels of social effects in FL.
As seen in Fig.~\ref{fig:simu12}, users usually obtain better payoffs given more participants under strong or weak social effects, as a larger network size can increase the chance of seeking better cooperating partners. Besides, the higher social effect also results in better individual payoff, validating the potential of social ties in FL.

\textbf{Incurred Overheads.}
{Fig.~\ref{fig:simu14} shows that the communication cost per iteration is in a decline and is less than $5$ KB, during the formation of a Nash-stable partition among $100$ participants. Besides, as a Nash-stable partition of social clusters is formed at $7$ iterations, no user tends to transfer to other social clusters for improved individual payoff, thereby the communication cost drops to zero after $8$ iterations.} Additionally, the social cluster formation process for each FL task only occurs at the initial phase of FL learning. Thereby, our SCFL only incurs small additional overheads when the social cluster changes. %by users to improve the availability of FL models.

\textbf{Convergence and Stability.} Fig.~\ref{fig:simu13} shows the social clustering results of our proposed FTU game in terms of the number of social clusters and average number of users per cluster.
In Fig.~\ref{fig:simu13}, the network starts with a random clustering partition with $40$ social clusters and converges to a final stable partition made up of $25$ social clusters with an average of $4$ users per cluster after $7$ iterations, which accords with the empty core and Nash-stability of our FTU game in Theorems~$2$--$3$. Notably, during $4$-$7$ iterations, there only exist inter-cluster member exchanges, i.e., no social users choose to form a singleton. As such, both the number of social clusters and average number of users per cluster remain unchanged during $4$-$7$ iterations.
Moreover, as seen in Figs.~\ref{fig:simu11} and \ref{fig:simu13}, the proposed Alg.~\ref{Algorithm1} can quickly converge to the Nash-stable partitions within only $7$ iterations {under FL when $N=100$}.

{\textbf{Discussions on Complex Models and Language Tasks.}
For different FL tasks and models, the fitting curve of the loss may vary.
However, the basic trend should remain the same, that is, with the increase of $\sigma$ or the decrease of $\gamma$, the accuracy of the model decreases and the loss increases. To validate this observation, we conduct additional experiments on the Resnet18 model for image classification tasks on CIFAR10 in Fig.~\ref{fig:simu4-4} and the {multilayer perceptron (MLP) model for text classification tasks} in Fig.~\ref{fig:simu4-5}.}

{First, we utilize the Resnet18 model to perform similar experiments on CIFAR10 with learning rate of $5\times 10^{-3}$, 30 participants, 30 global rounds, and local batch size of 64. As seen in Fig.~\ref{fig:simu4-4}, the loss cannot be fitted by 3D sigmoid curve, as the effects of $\gamma$ are more significant for the Resnet18 model on the CIFAR10 dataset than the CNN model on the MNIST dataset.
Instead, we observe that if $\gamma\leq1$, the effect of $\gamma$ is more obvious. When $\gamma > 1$, the effect of $\gamma$ is negligible and is masked by the effect of $\sigma$. The loss of the trained Resnet18 model can be fitted by a piecewise polynomial function:
\begin{align}
&\mathscr{L}\left( \sigma ,\gamma \right) =\nonumber \\
&~\resizebox{1.0\hsize}{!}{$
\begin{cases}
	1.951\!-\!2.132\gamma \!+\!14.21\sigma \!+\!1.163\gamma ^2\!+\!3.782\gamma \sigma \!-\!44.68\sigma ^2,\,\,\,&\gamma \!\leq\! 1,\\
	1.026\!-\!0.042\gamma \!+\!16.83\sigma \!+\!0.003\gamma ^2\!-\!0.0775\gamma \sigma \!-\!35.54\sigma ^2,&\gamma \!>\!1.\\
\end{cases}$}
\end{align}}\vspace{-4mm}

{Next, we train the MLP model using the 20newsgroup dataset\cite{lang1995newsweeder} for text classification task under FL with 5 clients, 10 global rounds, learning rate of 0.001 and local batch size of 32. As depicted in Fig.~\ref{fig:simu4-5}, unlike image classification tasks, the 3D fitting formula exhibits an exponential trend, where the loss continues to increase even when the noise disrupts the training process. The fitting formula is $\mathscr{L}= 2.053\exp(-0.0139\gamma + 1.1574\sigma) -0.7306$.}

\section{Conclusion}\label{sec:CONSLUSION}%\vspace{-0.118cm}
Striving a trade-off between privacy and utility plays a fundamental role in the practical deployment of DP-based FL services. Due to the rich manners of connectivity in real-world and motivations to ally for profits, we argue that the participants of FL tend to form social connections. By exploiting such social attributes among users, this paper has proposed a novel, efficient, and practical SCFL framework to realize feasible, high-utility, and customized privacy-preserving FL services.
Firstly, we have designed a contribution quantification and fair allocation mechanism for heterogeneous users in each social cluster. Considering some clusters may have low mutual trust, a customizable privacy preservation mechanism has been devised for adaptive local model update sanitization based on social trust. In addition, we have developed a distributed two-sided matching algorithm to obtain an optimized stable partition in FTU game. Experiment results have validated the effectiveness of SCFL in terms of user payoff, model utility, and privacy preservation, compared with existing solutions.

{In future work, we will further investigate the social-aware personalized FL framework by forming social clusters for users with similar social interests and learning a personalized model within each social cluster.}

\vspace{-4mm}
\bibliographystyle{IEEETran}
\bibliography{ref_SCFL}

\end{document}